\documentclass[acmsmall]{acmart}

\AtBeginDocument{%
  \providecommand\BibTeX{{%
    \normalfont B\kern-0.5em{\scshape i\kern-0.25em b}\kern-0.8em\TeX}}}

\setcopyright{acmcopyright}
\copyrightyear{2022}
\acmYear{2022}
\acmDOI{10.1145/1122445.1122456}

\acmJournal{TOSEM}
\acmVolume{1}
\acmNumber{1}
\acmMonth{6}

\usepackage[utf8]{inputenc}
\usepackage{array}
\usepackage{wrapfig}
\usepackage{multirow}
\usepackage{tabularx}
\usepackage{caption}
\usepackage{subcaption}
\usepackage{booktabs}
\usepackage{setspace}
\usepackage{tikz}
\usepackage{enumitem}
\usepackage{multirow}
\usepackage{longtable}
\usepackage{changepage}
\usepackage{hyperref}
\hypersetup{
    colorlinks=true,
    linkcolor=black,
    filecolor=black,      
    urlcolor=black,
    citecolor=black,
}
\usepackage{graphicx}
\usepackage{tasks}
\usepackage{fancybox}
\newcommand{\todo}[1]{}
\renewcommand{\todo}[1]{{\color{red} TODO: {#1}}}

\begin{document}

\title{The Influence of Human Aspects on  Requirements Engineering-related Activities: Software Practitioners' Perspective}

\author{Dulaji Hidellaarachchi}
\email{dulaji.hidellaarachchi@monash.edu}
\affiliation{%
  \institution{Faculty of Information Technology, Monash University}
  %\streetaddress{P.O. Box 1212}
  \city{Melbourne}
  \state{Victoria}
  \country{Australia}
}

\author{John Grundy}
\affiliation{%
  \institution{Faculty of Information Technology, Monash University}
  %\streetaddress{1 Th{\o}rv{\"a}ld Circle}
  \city{Melbourne}
  \country{Australia}}
\email{john.grundy@monash.edu}

\author{Rashina Hoda}
\affiliation{%
  \institution{Faculty of Information Technology, Monash University}
  \city{Melbourne}
  \country{Australia}}
 \email{rashina.hoda@monash.edu}

\author{Ingo Mueller}
\affiliation{%
 \  \institution{Faculty of Information Technology, Monash University}
  \city{Melbourne}
  \country{Australia}}
\email{ingo.mueller@monash.edu}


\begin{abstract}
Requirements Engineering (RE)-related activities require high collaboration between various roles in software engineering (SE), such as requirements engineers, stakeholders, developers, etc. Their demographics, views, understanding of technologies, working styles, communication and collaboration capabilities make RE highly human dependent. Identifying how \emph{"human aspects"}– such as motivation, domain knowledge, communication skills, personality, emotions, culture, etc.– might impact RE-related activities would help us improve RE and SE in general. This study aims to better understand current industry perspectives on the influence of human aspects on RE-related activities, specifically focusing on motivation and personality, by targeting software practitioners involved in RE-related activities. Our findings indicate that software practitioners consider motivation, domain knowledge, attitude, communication skills and personality as highly important human aspects when involved in RE-related activities. A set of factors were identified as software practitioners' key motivational factors when involved in RE-related activities, along with important personality characteristics to have when involved in RE. We also identified factors that made individuals less effective when involved in RE-related activities and obtained some feedback on measuring individuals' performance when involved in RE. The findings from our study suggest various areas needing more investigation, and we summarise a set of key recommendations for further research.
\end{abstract}

\begin{CCSXML}
<ccs2012>
 <concept>
  <concept_id>10010520.10010553.10010562</concept_id>
  <concept_desc>Computer systems organization~Embedded systems</concept_desc>
  <concept_significance>500</concept_significance>
 </concept>
 <concept>
  <concept_id>10010520.10010575.10010755</concept_id>
  <concept_desc>Computer systems organization~Redundancy</concept_desc>
  <concept_significance>300</concept_significance>
 </concept>
 <concept>
  <concept_id>10010520.10010553.10010554</concept_id>
  <concept_desc>Computer systems organization~Robotics</concept_desc>
  <concept_significance>100</concept_significance>
 </concept>
 <concept>
  <concept_id>10003033.10003083.10003095</concept_id>
  <concept_desc>Networks~Network reliability</concept_desc>
  <concept_significance>100</concept_significance>
 </concept>
</ccs2012>
\end{CCSXML}

\ccsdesc[500]{Software and its engineering~Requirements engineering}

\keywords{Human aspects, Requirements engineering, Software engineering}

\maketitle

\section{\textbf{Introduction}}
Requirements Engineering (RE)-related activities play a fundamental role in  software development, and are recognised as a critical part of software engineering (SE) to ensure fit-for-purpose and quality software products \cite{RN1601}.
RE-related activities include eliciting, analysing, documenting, validating and maintaining software requirements \cite{RN2732} \cite{RN2973}. According to Sommerville \cite{RN2974}: \emph{"RE is the process of establishing services that the customer requires from a system and the constraints under which it operates and is developed"}. In practice, RE is nowadays usually considered to be an iterative process in which the activities are interleaved with other SE tasks, and these RE-related activities are usually  carried out iteratively in methods such as agile software development or extreme programming \cite{RN2974} \cite{RN1644}. 
Like every other team activity, software engineering depends on individual performances and harmony among team members. According to Wick, there needs to be an effective collaboration of individuals with appropriate technical skills and complementary soft skills \cite{RN2733}. Hence, it is naturally challenging for  software developers and requirements engineers to work effectively together within the team and with external stakeholders who may have differences in terms of their human aspects.

In this work we define \emph{human aspects} in software engineering as human-related aspects that can become make-or-break issues in software projects \cite{RN1600}. Researchers have investigated various human aspects in various SE contexts -- such as \emph{personality} \cite{RN2956}\cite{RN2568}, \emph{emotions} \cite{RN2726}, \emph{motivation} \cite{RN2441}\cite{RN2618}, \emph{gender} \cite{RN2668}, \emph{culture} \cite{RN2593}, \emph{communication issues} \cite{RN2938}\cite{RN2383}, \emph{human errors} \cite{RN2967}, \emph{attitude} \cite{10.1145/1370114.1370127}, \emph{team climate} \cite{RN2712} and others -- and identified their impact on SE, for better or worse. 
We recently conducted a Systematic Literature Review (SLR) \cite{RN1600} on existing studies of human aspects impacting requirements engineering. In this study we  identified that human aspects, and their impact on RE-related activities, are still an area that has had relatively limited attention. The majority of the studies are academic-based that are conducted with academics and students and also focused on providing theoretical models/ strategies/ prototypes shows the need for more industry-focused studies. While \emph{motivation} and \emph{personality} have been studied in SE \cite{RN1614}, \cite{RN2576} \cite{RN2618}, we identified that these are two human aspects that to date have not been investigated much in relation to RE-related activities. When observing motivation and personality, we found many studies in the context of SE in general, with a few considering RE only as a part of it. However, we recognise that some studies have focused on the impact of motivation in RE combined with other human aspects, such as emotions, human values or personality. However, these studies have only paid attention to users, stakeholders or a particular RE activity e.g. requirements elicitation.

Hence, we wanted to better understand the  perspective of current industry practitioners on the influence of various human aspects on RE-related activities, to know whether our findings in our systematic analysis are in line with real-world experiences of software practitioners. In this study, we focus on identifying software practitioners' perspective on the influence of various human aspects when involved in RE-related activities, irrespective of the software development methodologies they follow, company size and domain of software applications they build. We were particularly interested in their views on (i) which human aspects they think impact RE-related activities, compared to those found in our SLR; (ii) what are their own key motivating factors when doing RE-related activities; (iii) which personality characteristics -- their own and their team mates -- do they think are most important when doing RE-related activities; (iv) can they identify any human aspects that can make RE-related activities less effective; and (v) how is their RE-related work performance measured and how do they think it should be measured. To do this, we conducted a survey study with the aim of understanding the industry perspective on the impact of human aspects on RE based on their experiences when involved in RE-related activities. The main contributions of this research are as follows:
\begin{itemize}
    \item Obtaining the perspectives of 111 current software practitioners involved in RE-related activities about the importance of various human aspects influencing them when they are involved in RE-related activities;
    \item Identifying a set of key factors that motivate individuals when involved in RE-related activities and their impact on achieving good RE/SE outcomes;
    \item Identifying a set of individual personality characteristics that practitioners think are important to have among software team members when conducting RE-related activities;
    \item Identifying a set of key factors that software practitioners think make them less effective when involved in RE-related activities and result in poorer RE outcomes;and
    \item A set of recommendations about further investigations needed to explore how various human aspects impact on RE-related activities, which would be beneficial for both academia and industry practitioners to measure and enhance RE-related work outcomes.
\end{itemize}

The rest of this paper is organized as follows. Section \ref{section 8} presents key related work in the research area with research gaps and, section \ref{section 2} introduces the research questions of this research. In  section \ref{section 3}, we have described the research methodology we used, including our survey design, procedures and how data were analysed. Findings are presented in section \ref{section 4}, and  section \ref{section 5} provides the discussion with key insights of the study and our recommendations. Finally, section \ref{section 7} presents the limitations and threats to the validity of the research and Section \ref{section 9} concludes this paper.

\section{\textbf{Related Work}} \label{section 8}
We summarise previous research that relates to our research area of this survey. Section \ref{section 8.1} describes studies related to human aspects in software engineering, section \ref{section 8.2} describes studies conducted related to human aspects in requirements engineering, and section \ref{section 8.3} describes studies related to performance assessment of software practitioners. 

\subsection{\textbf{Human aspects in Software Engineering}} \label{section 8.1}
A variety of human aspects have been shown to have an impact on different stages in the SE process. The majority of these studies have focused on their impacts during the software design and implementation stages. According to the systematic literature review (SLR) conducted by \cite{RN2966}, in these development stages,  designers and coders have been focused on in 94\% of their identified papers. A number of systematic studies have been conducted targeting the identification of various \textbf{human aspects of software engineers}, such as \textit{motivation, creativity, personality, behaviour, gender equity, human values, self-management barriers and self-compassion}. Cruz et al. \cite{CRUZ201594} in their systematic mapping study, reviewed  research on \textbf{personality} in SE. They analysed many published empirical and theoretical studies related to the role of personality on different aspects of SE. Based on their findings, pair programming, education, software engineers' personality characteristics, and team effectiveness related to personality were identified as the most focused on areas. 
\par Xia et al. \cite{RN3005} conducted a large scale study with software professionals to identify the relationship between project manager personality and team personality composition and project success. For this, they have investigated 28 completed software projects which contain 346 software professionals. To conduct this study, the DISC personality test was used and the researchers correlated the outcomes of the test with project success scores measured in six different dimensions: schedule, effort, risk, issue, quality, and customer satisfaction. The results indicate that project manager personality and team personality affect the success of software projects and suggest to focus on relationships between personality and software engineering activities as their study only demonstrates the link between personality and overall project success. Another empirical investigation was carried out by Kanij et al. \cite{10.5555/2819321.2819323} investigating the personality traits of software testers. They collected personality profiles of 182 software practitioners using IPIP test. 45.1\% of them were software testers and the majority of the rest were programmers. The results indicate that software testers are higher on conscientiousness factor than other software practitioners. 
\par Mendes et al. \cite{RN1602} conducted a study investigating the relationship between decision-making style and personality within the context of software project development where they conducted a survey and collected data from 63 software engineers. They identified seven statistically significant correlations between decision-making style and personality and built a regression model considering the decision-making style as the response variable and personality factors as independent variables. Vishnubhotla et al. \cite{RN3008} investigated the relationship between personality and team climate. They only focused on software professionals in agile teams in a telecom company. They used the FFM model for personality traits and the factors related to team climate (team vision, participative safety, support for innovation and task orientation) within the context of agile teams working in the telecom company. The findings indicated that there is a significant positive relationship between certain personality traits and team climate factors. The study also suggest to consider other human aspects in addition to personality traits to investigate their relationship between team climate. 

\par In \cite{RN2458}, a conceptual framework of programmer's \textbf{creativity} has been proposed and it incorporates personality traits of the programmers into this framework, as programmers' personality traits impact on their creativity intention. This framework is a theoretically established one based on three human aspects: personality traits, knowledge collection behaviour and creativity intention. After empirical analysis of the framework, the results are expected to support the theoretical base, which is considered human aspects positively impact on programmers' creativity. Lenberg et al. \cite{RN2968} conducted an SLR focusing on various \textbf{human behavioural} aspects in SE with the objective of creating a common platform for future research in the area. They suggested a new research area as \textit{behavioural software engineering} (BSE) and presented a definition of BSE as \emph{"the study of behavioural and social aspects of SE activities performed by individuals, groups or organisations"}. The results of their research indicated that BSE is an emerging research area where the majority of researches are based on software engineers, teams or organisations in general. They found that specific phases or activities in SE have not yet been frequently considered. Moreover, they identified that there is an imbalance of studies that focused on human aspects, as most of the studies considered \textit{communication, personality} and \textit{job satisfaction} related to software engineers. They suggest that researchers should explore more human aspects and consider their impact on different SE activities.

\par In \cite{RN2713}, an empirical study was conducted to investigate on how software testers can be \textbf{motivated}. Semi-structured in-depth interviews were conducted with 36 practitioners in 12 software organisations in Norway. Set of motivational and demotivational factors influencing software testing personnel were identified and proposed that combining testing responsibilities with variety of tasks engagement increase the satisfaction of testers which eventually increase their motivation. Sach et al. \cite{RN2582} also focused on motivational factors in software development where 23 software practitioners were engaged for a workshop on motivation and collected data to investigate motivational factors that affect on their software development practices. Based on their results, they claim the \emph{people} factor is the most commonly listed motivational factor for the software practitioners, compared to other factors such as financial, autonomy, learning and etc. In \cite{RN2572}, a systematic review was conducted to identify theory use in studies investigating the motivations of software engineers. By analysing 92 studies related to  motivation in SE, they found that many studies have focused on motivation of software engineers, but not explicitly underpinned by existing  motivational theories. However, the findings of the reviewed primary studies showed a clear relationship with these theories.   
\par \textbf{Gender} is another human aspect that is emerging in importance in SE research and practice. In \cite{RN2668}, a case study was used to investigate  gender equality in the SE context, a national software academy (NSA). This research tried to identify the experience of gender equality over three years in NSA and discussed  measures to be implemented in future research to raise awareness and reduce the gender gap among all levels at the NSA. In the study \cite{RN2682}, \textbf{human values} were measured related to SE where they investigated the influence of human values in the software production decision-making process. The researchers considered human values as a mental representation and investigated them based on three levels -- system level, personal level and instantiation level. Three human values prototypes were identified for software practitioner, the \textit{intrinsically-driven socially-concerned practitioner, the autonomous nonconforming risk-taker} and \textit{the fun-loving extrinsically-driven practitioner}. The researchers claimed that this approach should be used more widely so that researchers don't miss values in future research. A systematic mapping of \textbf{human cognitive biases} in software engineering was carried out\cite{mohanani2018cognitive}. This showed that software engineers are susceptible to a range of biased decision making at different phases of development. They highlight a lack of good mitigation techniques and limited theoretical foundations for interpreting biases in this area. They suggest some techniques to mitigate bias, but also highlight the need for further studies of biased human decision making in SE, including in RE.

\subsection{\textbf{Human aspects in Requirements Engineering}} \label{section 8.2}
Much of SE is in many aspects a human-centred activity \cite{john2005human}. In \cite{RN2952} \& \cite{RN2445}, it is argued that \textbf{Requirements Engineering} (RE) is arguably the most human-centred activity in SE, requiring people who are involved in RE needing to work closely and effectively with diverse stakeholders, software development team members, and other requirements engineers. In terms of studies that focused on the effects of human aspects on RE-related tasks in particular, \cite{RN2952}  focused on effective \textbf{communication} as a critical success factor during requirements elicitation. However, this study was limited to global software development (GSD) and identified that effective  communication plays a significant role in requirements elicitation specifically for GSD teams. It was found that \textit{geographical distribution}, \textit{time zone}, \textit{cultural diversity} and \textit{physical differences} were reasons for miscommunication when conducting requirements elicitation in GSD. Another SLR conducted in the GSD domain focused on identifying critical challenges in successful implementation of RE-related tasks \cite{RN2486}. Their analysis indicates that lack of effective communication, lack of knowledge sharing and awareness, lack of collaboration and organisational change are common critical challenges related to RE-related tasks in GSD. In their study \cite{RN2215}, Khan and Akbar performed an SLR and an empirical investigation on \textbf{motivation factors} for the requirements change management process in GSD. They explored the motivators that contribute to requirements change management by extracting 25 motivators and finally developed taxonomies of identified motivators such as accountability,  clear change management strategy, overseas site's response, effective requirement change management leadership, etc. Aldave et al. \cite{RN2969} conducted an SLR to identify the influence of \textbf{creativity} on requirements elicitation within agile software development. They found that enhancing creativity in requirements elicitation can be implemented successfully in agile based software projects, specifically, user interface development projects. Moreover, they identified that creativity is an important aspect in SE which brings innovation to the project. Despite their findings, they say that more research is required to understand the influence of creativity in RE, as their study was limited to just the requirements elicitation phase in agile software development. In \cite{RN2430}, it is focused on classifying effective \textbf{personalities} for web development in requirements elicitation. Their research revealed that there is a relationship between human personalities and RE in web development and need more research that consider more human aspects and their impact on RE-related activities. 
\par Most of these systematic and empirical studies have  focused on various human aspects related to SE in general, or predominantly design, development, agile teams and GSD contexts. The studies that focused on RE have mainly been limited to GSD, web development domain or a particular activity in the RE; usually requirements elicitation. Cheng and Atlee \cite{RN1645} discussed current and future research directions in RE. They claim that identifying human behaviour in RE is an open and very challenging problem and it has become a key emerging area for RE researchers. As longer-term actions that would help the RE research community, they state that RE researchers should think beyond current RE and SE knowledge and collaborate with other disciplines to improve RE-related activities, including identification of better methods to model human behaviours in RE. 

\subsection{\textbf{Performance Assessment of Software Practitioners}} \label{section 8.3}
\textcolor{black}{Measuring performance of individuals by examining their work based on a set of pre-defined criteria and providing feedback for improvement are important parts of many organisational environments. For example, work performance in requirements engineering can be defined as behaviours that are relevant to accomplish a RE certain task in a specific situation \cite{RN1646}.} There are various criteria/models that have been used to assess individual performance and these vary based on the organisation or the individuals they assess. Killingsworth et al. \cite{RN1639} presented a model to motivate and evaluate information systems staff and the model consists of five factors; \emph{product quality}- employee's contribution to the delivered product or service quality, \emph{staff development}- employee's contribution to staff development, including personal development of each individual as well as building high-performance teams, \emph{customer outreach}-employee's contribution to maintain and expand work with current clients as well as the effort to win new clients, \emph{administrative efficiency}- employee's contribution to develop and maintain administrative procedures with punctuality and accuracy, and \emph{fiscal responsibility}-employee's financial plan evaluated against organisation's financial plan for the employee. In this model, a senior project manager and team leader are the people who will assess each employee against each of these factors with the use of five point rating scale. There are some other approaches presented in various studies and most of them are specifically for programmers' performance assessment.
\par In \cite{10.1145/1142635.1142640}, \emph{Programmer Appraisal Instrument (PAI)} has been introduced where programmers are reviewed on four major areas and total 42 questions are used to assess the performance on those areas. Berger and Wilson's \cite{10.1145/1142620.1142629} \emph{Basic Programmer Knowledge Test (BPKT)} is another performance assessment model which is specifically designed for programmers. This test evaluates the knowledge of the programmer based on six areas; logic estimation and analysis, flow diagramming, programming constraints, coding operations, program testing and checking and documentation and these areas will be assessed with 100 multiple choice questions. In the study \cite{10.1145/800115.803716}, 13 categories are presented to assess the performances of programmers and analysts, and a five point scale will be used to rate each category. For each category, there are set of questions which is used to define the categories and do the appraisal. Upon completion of the evaluation, each programmer will be given a private review, and guidelines for self-improvement.

%\begin{itemize}
%\item \textit{\textcolor{black}{Performance Assessment in Requirements Engineering}}
%\end{itemize}

Considering performance assessment in requirements engineering, not much published research is available. Regarding performance in RE, there are three dimensions that have been considered, namely, \textbf{quality of RE service}, \textbf{quality of RE products} and \textbf{RE process control}. The quality of RE service refers to the service that is provided to the users during requirements engineering and it is measured using two criteria; \emph{(1) user satisfaction and commitment and \emph{(2) fit of recommended solution with the organisation} \cite{RN1612} \cite{RN1638}}. The quality of the RE product is measured with two significant product qualities; \emph{(1) quality of architecture} and \emph{(2) quality of cost/benefit analysis}. Based on the stakeholders' ratings, it was identified that the quality of RE service  is the most important one when measuring RE performance and they have considered "good interaction between all groups" when providing their ratings. The quality of RE product was rated as the second important dimension when measuring the performance while the process control, the third dimension rated as the lowest which is related to variations of the cost, duration and effort of the process. However, these performance measurement dimensions are rated by the stakeholders and have not been assessed with managers and software practitioners involved in RE-related activities. Hence, it is important to identify the dominant factors related to performance assessment in RE from the perspective of software practitioners involved in RE-related activities. This highlights the importance of identifying the software practitioners' perspective on these areas who are actually involved in RE-related activities. This motivated us to design and conduct this survey study with the aim of obtaining answers to the following research questions.

\section{\textbf{Research Questions}} \label{section 2}
Our main objective for this research was to investigate the current industry perspective on the impact of various human aspects on RE-related activities. We wanted to find out what human aspects they think impact on performing RE-related activities. We wanted to find out what things motivate them when performing RE-related activities.  We wanted to find out which personality characteristics they think are important for themselves and teammates that impact RE-related activities. We wanted to find out which factors make individuals less effective when involved in RE-related activities and we wanted to find out how the performance of RE-related work can be measured.
To achieve these objectives we  wanted to find answers to the following research questions:                                     
%\color{black}
% \cornersize{.2} 

%\ovalbox{ \centering\begin{minipage}{38em}

\par \textbf{RQ1. What human aspects influence individuals when performing RE-related activities?} \textcolor{black}{This research question focuses on understanding industry perspectives of the importance of  a set of candidate human aspects that we identified from our analysis of the literature as impacting on RE-related activities via our SLR \cite{RN1600}, and how these aspects are perceived as influenced the performance of RE-related activities by software practitioners.} \\
%\end{minipage}}

%\vspace{0.5cm}

%\color{black}
% \cornersize{.2} 
%\ovalbox{\begin{minipage}{38em}
\par \textcolor{black}{\textbf{RQ2. What factors motivate individuals to  perform effectively their RE-related activities?} From our SLR, we identified ``motivation" as one of the key human aspects that has been less investigated in relation to performing RE-related activities. As effectiveness refers to the quality of desired results \cite{article}, this research question focuses on identifying software practitioners' opinions on motivation, what affects their motivation, and the impact of these factors on their effectiveness when performing  RE-related activities. We used a set of high-level factors that have been reported related to individuals' performance in RE in our survey  \cite{RN1612} \cite{RN1613} \cite{RN1646}.\\}
%\end{minipage}} 

 %\color{black}
% \cornersize{.2} 
%\ovalbox{\begin{minipage}{38em}
\par \textcolor{black}{\textbf{RQ3. What personality characteristics are important to perform effectively their RE-related activities?} From our SLR, we identified ``personality" as one of the key human aspects that has been less well investigated in relation to performing RE-related activities. This research question focuses on identifying the  personality characteristics that our respondents think are important when performing RE-related activities, by considering both an individual's personality and the personality characteristics they would like to see in their team mates.\\ }
% \end{minipage}} 

% \vspace{0.5cm}

% \color{black}
% \cornersize{.2} 
%\ovalbox{\begin{minipage}{38em}
\par \textcolor{black}{\textbf{RQ4. What factors make individuals less effective when performing RE-related activities?} We wanted to identify any motivational, personality or other factors that make individuals less effective when performing RE-related activities. \\}
 %\end{minipage}} 
 
 %\vspace{0.5cm}

 %\vspace{0.5cm}
 
% \color{black}
% \cornersize{.2} 
%\ovalbox{\begin{minipage}{38em}
\par \textcolor{black}{\textbf{RQ5. What factors could be used to measure an individual's performance when involved in RE-related activities?} Respondents gave us their opinions on the importance of human aspects when involved in RE-related activities that impact their RE performance. But what does it mean to perform well on RE-related activities? There is limited research on how software practitioners' performance is can be measured when involved in RE-related activities. We wanted to get software practitioners feedback on how RE-related work performance is currently measured and their views on how it should be measured, including human aspects and technical aspects of RE-related work}.
% \end{minipage}} 

\section{\textbf{Research Methodology}} \label{section 3}
We employed a survey research method to conduct this study, as we wanted to reach out to a broader population of software practitioners to know their perspectives on the influence of human aspects on RE-related activities \cite{RN1599} \cite{RN1598}. 

\subsection{Survey Design} \label{section 3.1}

%\todo{Put a PDF not the actual survey as link - easier for reader/referee:} - \textcolor{black}{added the questions in appendix}

\begin{figure}[]
  \includegraphics[width=\linewidth]{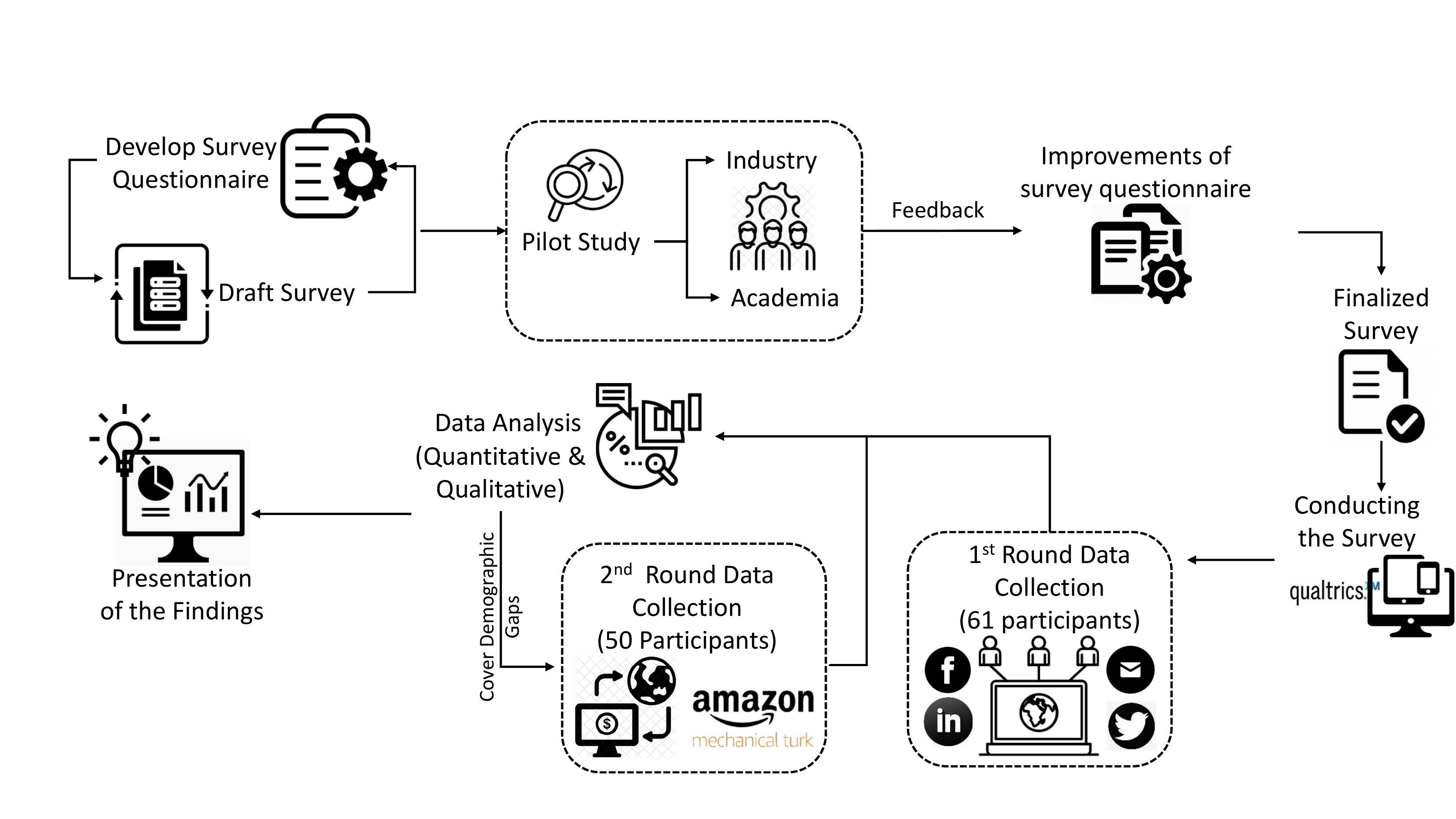}
  \caption{Overview of the research methodology of the study }
  \label{Methodology figure}
\end{figure}

Our survey\footnote{\url{https://figshare.com/s/ca2c5098c1ad4de34c13}} was designed to get an overview of participants' perspectives on the influence of various human aspects when involved in RE-related activities based on their experiences. To achieve the aim of this study, the survey was designed following each step as shown in figure \ref{Methodology figure}. The planning of conducting this survey was performed between October 2020 and December 2020. During this phase, we defined the survey goals, variables and designed the questionnaire following several iterative processes prioritizing the essential questions to include in the survey. Hence, the survey was designed with both closed and open-ended questions where for most of the closed-ended questions, a Likert scale with five possible responses ("Not at all Important" to "Extremely Important") were used. The survey consists of 22 questions, split into four main sections as follows focusing on several areas in-lined with our future studies (Appendix \ref{A}). 

\subsubsection{Participants Information} \label{section 3.1.1}
The first and second sections of the survey were designed to collect \textbf{basic} demographic information and the current employment details of the participants. The survey was anonymous and did not record any personally identifiable details. We only elicited  information on participant gender, age ranges, country of residence and educational qualifications. With regards to their current employment details, we collected their job role/title and summary of their main job responsibilities to identify whether they are involved in RE-related activities. We asked about the level of experience they have and the type of software development methods they involved in where they can select from given options (traditional, agile or both) and mention the exact methodologies they use. We wanted to collect and analyse participant opinion in general about influence of human aspects on RE-related activities, without trying to correlate to e.g. company or team size, particular development method, particular software domain etc.% \textcolor{black}{The domain of RE done (e.g. bespoke software vs. market-driven software) was not taken into consideration. How teams in each do different aspects of RE, as well as the potential of different human aspects to influence and being influenced by these contextual factors, is potentially large. This will require future studies to further find and examine any similarities and differences. }

\subsubsection{Understanding the Participants' Perspectives of Human Aspects} The third and fourth sections were designed to get an overview of the participants' perspectives of the influence of human aspects on RE-related activities as well as to get a detailed idea of their views on the particular human aspects of motivation and personality. In section three, we  focused on identifying what human aspects they considered most important in terms of their influence on RE-related activities. The participants were asked to rate the importance of human aspects based on their experience. We provided a list of human aspects that were identified in our previous study \cite{RN1600} and asked them to provide any other human aspect(s) via an open-ended question.

\textcolor{black}{We then wanted to get more details on their views of assessment of performance of RE tasks, motivation, and personality, all identified in our recent SLR as under-researched areas for RE-related activities that we were particularly interested in. Section three consists of questions to identify factors that they think are important in measuring performance when involved in RE-related activities and factors that motivate software practitioners to perform RE-related tasks. 
We also asked them to identify any factors that they think make them less effective in performing RE-related activities.}
\textcolor{black}{According to the literature, performance can be described as all behaviours that are relevant to accomplish a certain task in a specific situation and it is important to know how to measure software practitioners' performance when involved in RE-related activities \cite{RN1612}. Though there is currently limited research available on measuring performance of software practitioners involved in RE-related activities, we used a set of high-level factors that have been reported related to individuals' performance in RE in our survey  \cite{RN1612} \cite{RN1613} \cite{RN1646}, and we asked participants to rate their importance according to participants' opinions}. 

Section four focused on eliciting participants' experiences on how individual characteristics related to personality are important when conducting RE-related activities. We wanted to identify their experiences considering their individual personality characteristics as well as their team members' personality characteristics. The personality characteristics listed were based on the well-known "Five Factor Model" of personality, which is one of the most popular models to describe individual characteristics (personality), grouping them into five broad dimensions, Extraversion, Agreeableness, Conscientiousness, Neuroticism and Openness to Experience \cite{RN1603} \cite{RN1604}. %As our next stage of the research is planned to focus on the influence of personality and motivation on RE-related activities, at the end of the survey, we asked their willingness to participate in the next stage, and if they are interested in participating, we asked to provide their contact details (name and email address), which were kept separately prior to the data analysis.  

\subsubsection{Pilot Study} After designing the survey, we conducted a pilot study with software practitioners from our networks to verify the clarity and understandability of the questions, the time reported to complete the survey and to get their suggestions on improving the survey. The survey was sent to three software practitioners who are currently involved in RE-related activities and three academics who have prior experiences in software industry performing RE-related activities. All of them provided feedback on the survey questions and based on that, we modified the questions by rewording and adding definitions to some questions to make them more clear and understandable. With the improvements, we finalized the survey and conducted our main study.

\subsection{Survey Sampling and Data Collection} \label{section 3.2}
The target population of our survey is software practitioners involved in RE-related activities, and therefore, we used a non-probabilistic purposive sampling technique in the study as our selection is based on specific characteristics other than availability (as our target participants are software practitioners involved in RE-related activities) \cite{RN1643}. We used the \emph{Qualtrics} platform to design the survey and after obtaining the required ethics approval (Reference Number: 26219), we advertised the survey as an anonymous survey link. 

Two rounds of data collection were carried out and the first round of data collection was performed between December 2020 and February 2021, where the survey was advertised in social media groups (LinkedIn, Twitter and Facebook) focusing on software practitioners involved in RE-related activities, where 61 practitioners completed the survey (results from industry practitioners). After the first round of data collection, we wanted to try and cover the gaps in demographic information of the voluntary first round participant pool and to that, we conducted a second round data collection. Similar to the study \cite{RN1611}, we advertised our survey in the Amazon Mechanical Turk (AMT) between March 2021 to April 2021, where a total of 50 software practitioners completed the survey with useful results. The Amazon Mechanical Turk platform has built-in options to filter participants for participant selection and customize a monetary incentive for each participant. Particularly, we applied the participant filter options of “Employment Industry - Software \& IT Services” and “Job Function - Information Technology” to ensure that we reached our desired target population. The participants who completed the survey via AMT were given a reward of 6.40 AUD after completing the survey. As the survey was shared worldwide, we received responses from many countries, as shown in Table \ref{TABLE 1: Participants' demographics}. Due to our professional networks, the majority (42\%) came from Sri Lanka, and a significant number of others were from the USA, Brazil, Italy, etc. The detailed analysis of the participants demographics are provided in section \ref{section 4.1}.\textcolor{black}{ We didn't ask details about particular domain of work in our survey. Thus, we did not try and find participants in diverse domains in our recruitment process.
}

\begin{table}[t]
\centering
\caption{\centering Data sources and analysis types used to answer RQs (statistical analysis for quantitative data analysis and STGT for qualitative data analysis \cite{RN1609})}
\label{TABLE: Data analysis methods}
\resizebox{0.9\linewidth}{!}{%
\begin{tabular}{@{}llll@{}}
\toprule
\multicolumn{1}{l}{\textbf{RQ} }      & \textbf{Data Source} & \textbf{Data Analysis Type}    & \textbf{Purpose of analysis}                                            \\ \midrule
\multirow{2}{*}{\begin{tabular}[c]{@{}l@{}}  RQ1\end{tabular}}  & {\begin{tabular}[c]{@{}l@{}}  Closed-ended question \end{tabular}} 
 & Quantitative analysis & \multirow{2}{*}{\begin{tabular}[c]{@{}l@{}}To get an overview of software practitioners' perspective on the \\influence of human aspects when involved in RE-related activities  \end{tabular}}\\
 & {\begin{tabular}[c]{@{}l@{}}Follow-up open-ended \\question\end{tabular}} & Qualitative analysis \\

RQ2 & Open-ended Question & Qualitative analysis & {\begin{tabular}[c]{@{}l@{}}To identify what factors motivate software practitioners to\\ perform effectively RE-related activities and their impact\end{tabular}}\\
{\begin{tabular}[c]{@{}l@{}}  RQ3 \end{tabular}}  & {\begin{tabular}[c]{@{}l@{}} Closed-ended question \end{tabular}} 
 & Quantitative analysis  &{\begin{tabular}[c]{@{}l@{}} To identify what personality characteristics are important to \\perform effectively RE-related activities (individual as well \\as their team)  \end{tabular}}  \\
 
 RQ4  & Open-ended question & Qualitative analysis & {\begin{tabular}[c]{@{}l@{}} To identify factors that make software practitioners less effective when \\performing RE-related activities \end{tabular}} \\
RQ5 & Closed-ended question & Quantitative analysis & {\begin{tabular}[c]{@{}l@{}} To get an idea of software practitioners' views on measuring performance\\ when involved in RE-related activities \end{tabular}} \\

\\

\bottomrule
\end{tabular}%
}
\end{table}
\subsection{Data Analysis} \label{section 3.3}

\begin{figure}[]
  \includegraphics[width=\linewidth]{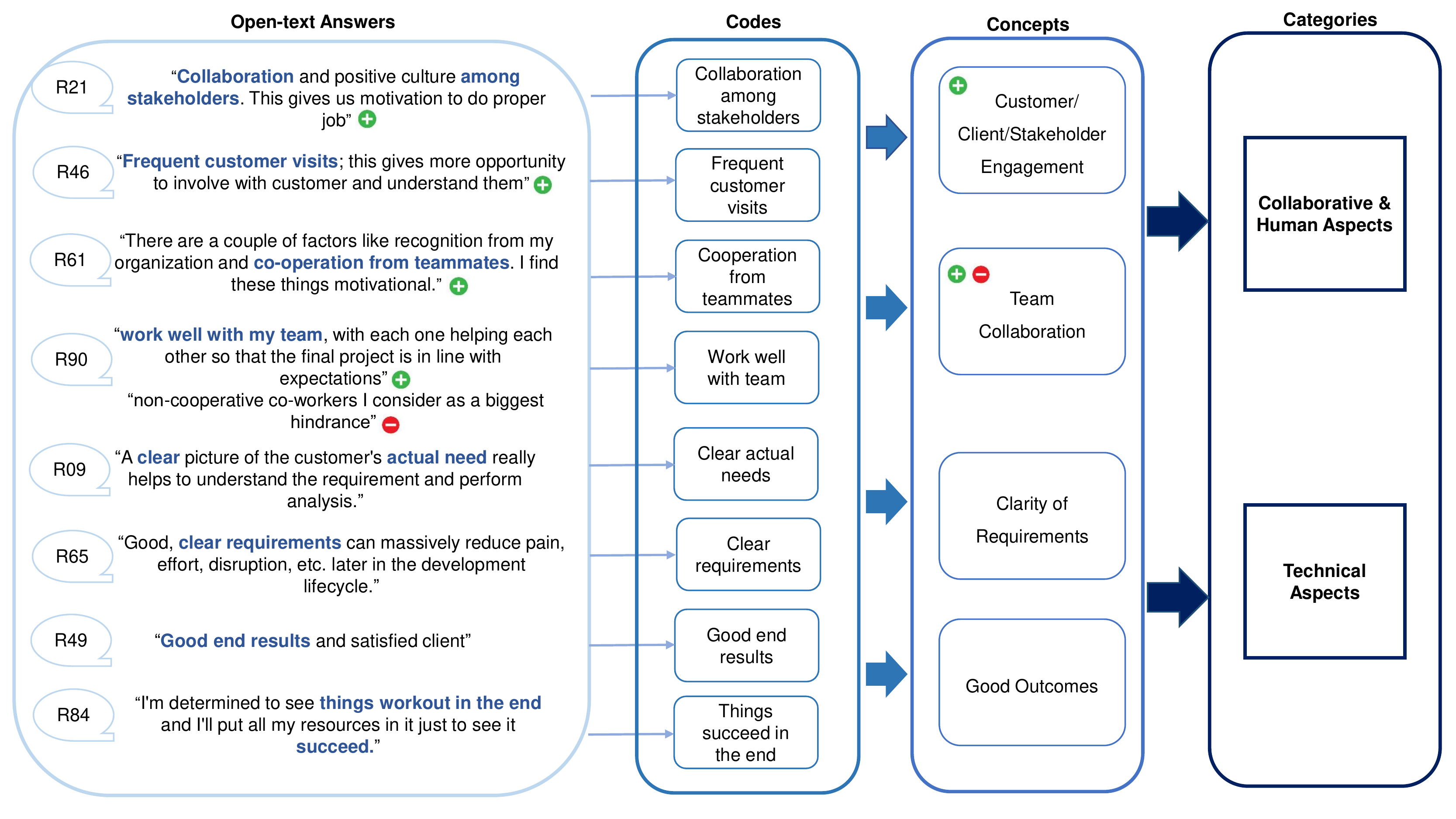}
  \caption{Basic data analysis steps of the Socio-Technical Ground Theory (STGT) \textit{for Data Analysis} \cite{RN1609} followed for qualitative data analysis}
  \label{Fig 7: STGT steps}
\end{figure}

Both qualitative and quantitative data was collected from our survey. Therefore, we used a mixed method approach for the data analysis of this study and Table \ref{TABLE: Data analysis methods} shows the data analysis types (quantitative/qualitative) we used to answer each research question of this study. Statistical analysis was conducted on the quantitative data using Microsoft Excel. Excel was also used to organise the qualitative data. For the qualitative data analysis, we used Socio-Technical Grounded Theory (STGT) \textit{for Data Analysis} \cite{RN1609}, which is particularly suited for analysing qualitative data, such as that collected from open-ended text-based survey responses (or \textit{open-text} for short). Unlike a \textit{full STGT study}, that typically requires extensive qualitative data, \textit{STGT for data analysis} does not aim for advanced theory development. Instead, it aims to identify key patterns in the qualitative data and presents them as layered and/or multi-dimensional findings along with insights and reflections, explained below. To do this, we followed an \textit{open coding} approach to generate concepts and categories with \textit{constant comparison} of various open-text answers. For example, for the question \emph{"In your opinion, what are the factors that motivate you to perform effectively in requirements engineering activities? Please explain briefly why"}, we received open text answers from 102 participants that can be analysed using the STGT \textit{for data analysis} approach.  As shown in figure \ref{Fig 7: STGT steps}, open coding was applied in open-text answers and codes such as \emph{collaboration among stakeholders, frequent customer visits, engage with stakeholders} were combined to create the concept \textbf{\emph{customer/client/stakeholder engagement}}, while \emph{collaborative work}, \emph{engage with co-workers}, \emph{cooperation among team} were combined to create the concept \textbf{\emph{team collaboration}}. Codes such as \emph{clear requirements, clarity of requirements, and well-thought-out requirements} led to the concept \textbf{\emph{clarity of requirements}}. 

Likewise, we used open coding for each open-text answer in the survey, and with constant comparison we grouped the codes into various concepts. For the above-mentioned question, the set of concepts was identified as factors that motivate individuals when performing RE-related activities. Moreover, based on the explanations given by the participants in the open-text answers, these concepts were identified as positive (denoted by the icon \includegraphics[height=1em]{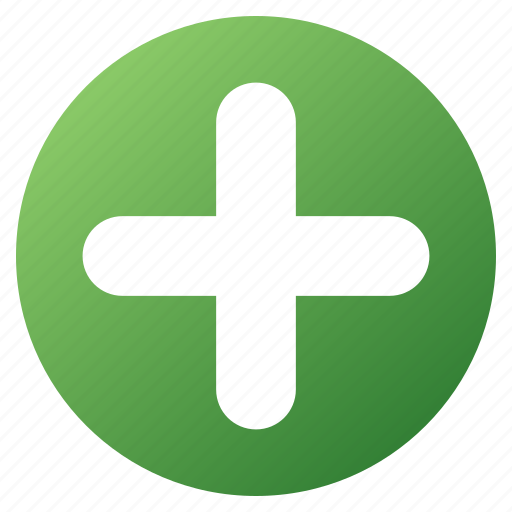})  and negative (denoted by the icon \includegraphics[height=1em]{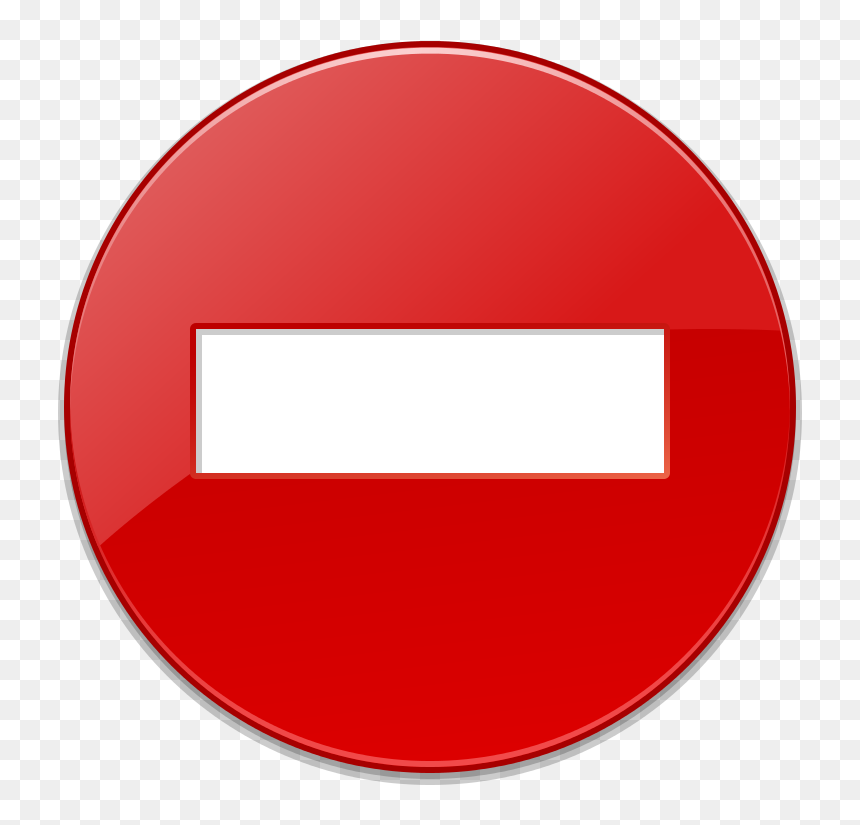}) where we were able to identify positively and/or negatively impacted motivation factors. Then the related concepts were combined to derive categories. This resulted in two high level categories namely; \textbf{collaborative \& human aspects} and \textbf{technical aspects} that affect individuals motivation when involved in RE-related activities. 

\par All of the authors were involved in each stage of the survey from designing the draft questionnaire to presentation of the findings. Several discussion rounds were conducted to make decisions for each stage. In the analysis stage, quantitative data was analysed by the first author and shared among the others to discuss each step followed, and identify best ways to present the findings. The third author conducted initial training sessions for the team on using the STGT approach for qualitative data analysis and was involved in the qualitative data analysis process with the first author. Over the course of the project, the team held several discussion rounds to finalize the identified concepts and categories based on the qualitative data and contributed to presenting the findings.

%\begin{itemize}
%\item \textit{Role of Memos}
%\end{itemize}
Using the STGT \textit{for data analysis} involves open coding, constant comparison, and memoing. ``\textit{Basic memoing is the process of documenting the researcher’s thoughts, ideas, and reflections on emerging concepts and (sub)categories and evidence-based conjectures on possible links between them}'' \cite{RN1609}. Thus, we wrote \textbf{memos} to record key insights and reflections generated while following the open coding activities. Below is an example of a memo that we recorded related to communication as a motivation factor. We have discussed the key \textit{insights and reflections} generated from memoing in section \ref{section 5.2}. \\

 \color{black}
 \cornersize{.2} 
\ovalbox{\centering \begin{minipage}{36em}

\par \textbf{Memo on ``Communication as a Motivation Factor''} Some participants (n=10) mentioned \textit{communication} as an motivation factor for them to perform RE-related activities. They have mentioned it as \emph{"good communication"}, \emph{"friendly communication"}, \emph{"proper communication"} or \emph{"direct communication"} which provides a reasonable understanding of the importance of positively-oriented communication in motivating practitioners to perform their RE-related activities. These can be considered desirable characteristics or attributes of communication that can positively influence motivation. It is also interesting to note that some of these adjectives are subjective and need more details to understand them better. For example, ``friendly'' communication can denote both the tone and the approach of communication. Also, what is ``friendly'' to one practitioner may not be for another. Understandably, this was not further explained by the participants in their relatively limited open-text survey responses. They have also made a connection between communication and who it is happening with by relating it to clients/customers (e.g., clarity of conversation with clients). These preliminary relationships can be further investigated in future studies focused on communication aspects. 
\end{minipage}}

\section{\textbf{Findings}} \label{section 4}
We present the key findings from our survey analysis related to each research question. Survey questions can be found in Appendix \ref{A}.

\begin{table}[t]
\centering
\caption{\centering Participants' Demographics Information}
\label{TABLE 1: Participants' demographics}
\resizebox{0.8\columnwidth}{!}{%
\begin{tabular}{@{}llll@{}}
\toprule
\multicolumn{1}{l}{\emph{Demographic Information} }      & \emph{Overall \% } & \emph{1st Round } & \emph{2nd Round }                                                \\ \midrule
\multicolumn{1}{l}{\textbf{Education Information of the Participants}}                                                        \\ \midrule
University degree in Software engineering/ Computer science     & 58.5\% & 29.7\% & 28.8\% \\
University degree in other IT field     &  21.6\%  & 18\% & 3.6\%   \\
Associate degree/ diploma in Software engineering/ Computer science    &  5.4\%   & - & 5.4\%          \\
Associate degree/ diploma in other IT fields                 &  1.8\%      & - & 1.8\%       \\
Other university degree/ associate degree/ diploma       & 12.6\%  & 7.2\% & 5.4\%          \\ \midrule
\multicolumn{1}{l}{\textbf{ Job roles of the Participants}}   \\ \midrule
Software Engineer                                                & 28.8\%   & 8.1\% & 20.7\%        \\
Business Analysts                       & 18.0\%  & 13.5\% & 4.5\%  \\  
Senior Software Engineer & 9.0\%          & 8.1\% & 0.9\%                                                      \\
IT project Manager & 5.4\% & 3.6\% & 1.8\% \\
Senior Business Analysts         & 4.5\%  & 3.6\% & 0.9\% \\
 IT Consultant                 & 4.5\% & 2.7\% & 1.8\%\\
 Senior Quality Engineer      & 4.5\% & 4.5\% & - \\
Software QA Engineer  & 3.6\% & 2.7\% & 0.9\%
\\
 IT System Administration \& Development & 3.6\% & - & 3.6\%
\\
Senior Consultant, Product owner & 2.7\% each & 2.7\% each & -
\\
 Software Architect  & 2.7\% & 0.9\% & 1.8\%
\\
 IT Specialist  & 2.7\% & - & 2.7\%
\\
 Tech Lead & 2.7\% & 1.8\% & 0.9\%        
\\
IT Director, Computer Scientist & 1.8\% each & - & 1.8\% each \\
IT intern & 0.9\% & - & 0.9\%\\ \midrule
\multicolumn{1}{l}{\textbf{Countries of the Participants}}   \\ \midrule
Sri Lanka                                           & 42\%  & 42\% & - \\
USA    & 18\%  & 3\% & 15\% \\
Brazil   & 9\%  & - & 9\%         \\
Italy              & 8\%  & - & 8\%            \\
Canada                                  & 3.6\%       & - & 3.6\%              \\
Sweden       & 2.7\%   & 2.7\% & -                                              \\
Australia, Indonesia         & 1.8\% each     & 1.8\% each & -                                                                  \\
UK, France                  & 1.8\% each   & - & 1.8\% each                                                                    \\
\begin{tabular}[c]{@{}l@{}}Saudi Arabia, Suriname, Singapore, Austria \end{tabular}                                         & 0.9\% each & 0.9\% each & -  \\  
\begin{tabular}[c]{@{}l@{}} Thailand, Kenya, Germany, Spain, Albania, Ecuador \end{tabular}                                         & 0.9\% each & - & 0.9\% each  \\
\bottomrule
\end{tabular}%
}
\end{table}

\subsection{Participants' Demographics} \label{section 4.1}

In total 111 software practitioners participated in the survey.  The majority were male (63.1\%), with ages ranging between 26 to 35 years (56.8\%). The most common roles are software engineer (28.8\%) and business analyst (18.0\%), and the most commonly repeated education level is a university degree in software engineering or computer science (58.5\%). Table \ref{TABLE 1: Participants' demographics}  summarises these overall statistics as well as provides a comparison of participants' demographics between first and second round of data collection. In the first round, majority were business analysts (13.5\%) and in the second round, the majority were software engineers (20.7\%). The major geographic information gap was related to the countries of the participants where we got 42\% of responses from Sri Lanka. To reduced the biasness, we restricted responses from Sri Lanka in the second round data collection that we conducted using AMT.  There, we obtained considerable amount of responses from USA (15\%), Brazil (9\%) and Italy (8\%). Most of the participants have 1 to 5 years of work experience in the software industry (54\% of participants). \textcolor{black}{The majority (45.9\%) use agile software development methods, such as scrum, kanban, XP, crystal, and combinations such as scrumban. 37.8\% of our survey participants said they follow more traditional (waterfall) software development methods, and 15.3\% use both agile and traditional methods. }
\begin{figure}[]
\centering
\begin{minipage}{.5\textwidth}
  \centering
  \includegraphics[width=\linewidth]{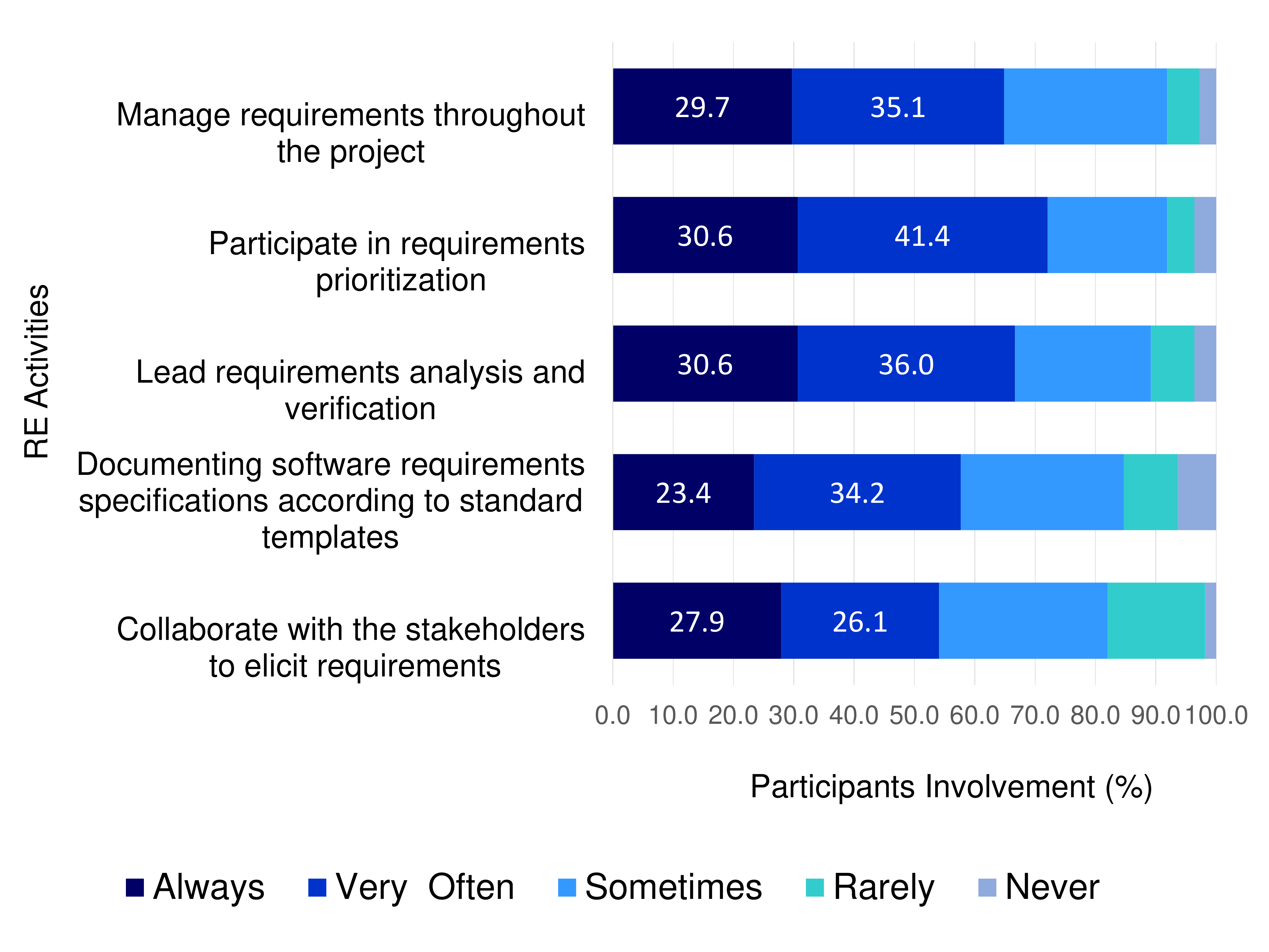}
  \captionof{figure}{Participants' involvement in requirements \\engineering activities}
  \label{Fig 2: RE-related activities}
\end{minipage}%
\begin{minipage}{.55\textwidth}
  \centering
  \includegraphics[width=\linewidth]{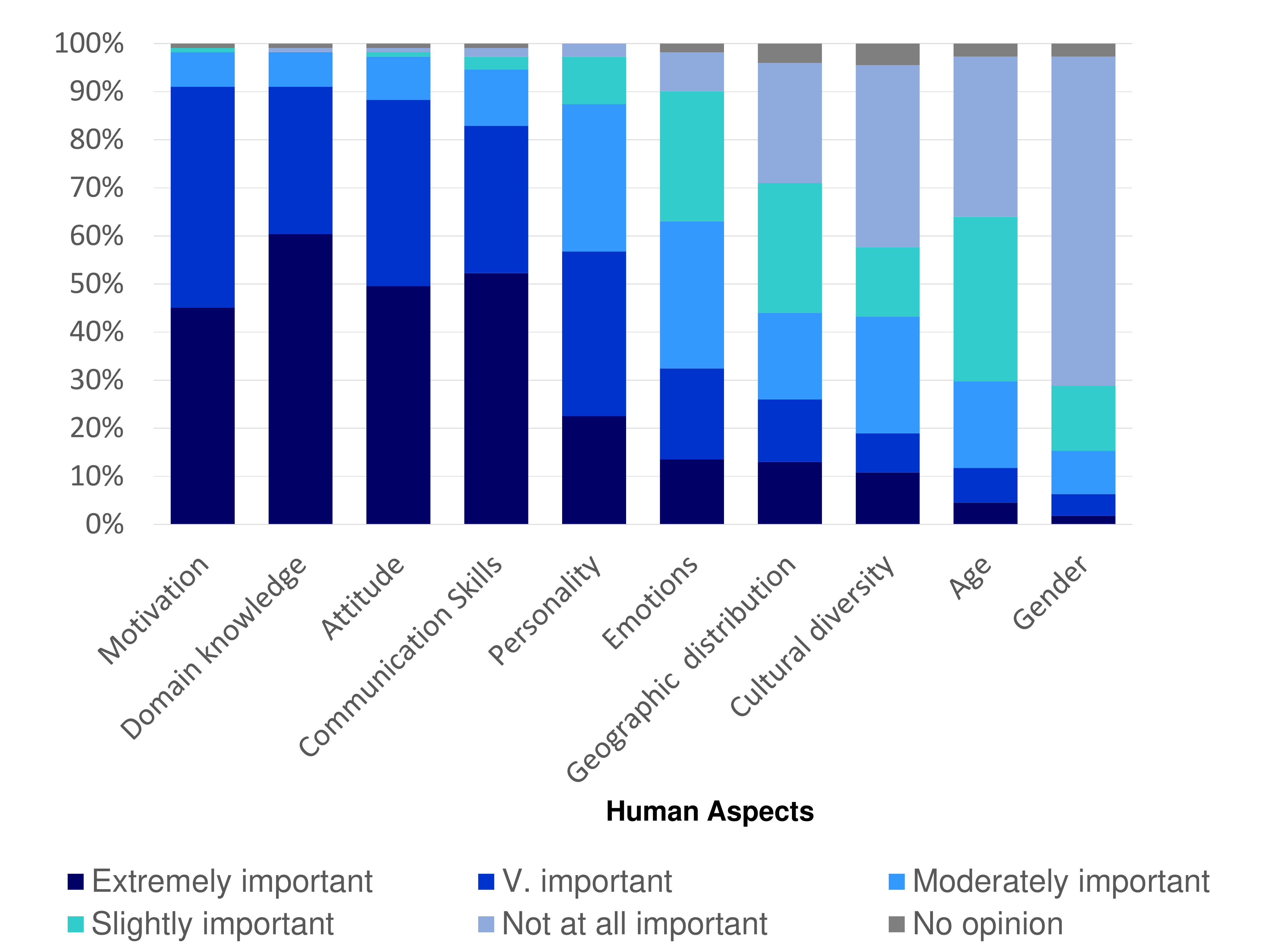}
  \captionof{figure}{Level of influence of the human aspects on the performance of individuals in RE-related activities: Practitioners' perspective}
  \label{Fig 1: Human aspects in RE}
\end{minipage}
\end{figure}

\par \textcolor{black}{As our target survey participants were  software practitioners involved in RE-related activities, we asked  participants to state their level of involvement in major RE-related activities that we provided in the survey. As shown in figure \ref{Fig 2: RE-related activities}, the majority of our participants are either always or very often involved in major RE-related activities. These activities include \emph{collaboration with stakeholders to elicit requirements, documenting software requirements specifications according to standard templates, lead requirements analysis and verification, participate in requirements prioritization and manage requirements throughout the project}. From our survey participants' answers, the majority are involved in requirements prioritization (72\%), lead requirements analysis and verification (66.7\%) and manage requirements throughout the project (64.8\%). 57.7\% are involved in documenting software requirements specifications, and 54\% always or very often collaborate with stakeholders to elicit requirements. Only 3.6\% of participants mentioned that they have never been involved in one of these RE-related activities. Apart from the given RE-related activities list, we also asked our  participants to mention any other key job responsibilities they have. 34 of 111 reported these other job responsibilities. Among them, the majority (12 out of 34) mentioned testing of user requirements (35.3\%) and designing the solution/prototyping (35.3\%) where the others (10 of 34) mentioned team and customer management. These results indicate that we have achieved our aim of collecting data from our target participant group irrespective of the domain/context in which they work.} 

\subsection{RQ1 -- Participants' perspective on the influence of human aspects on RE-related activities} \label{section 4.2}
In the systematic literature review we conducted \cite{RN1600}, we identified a set of human aspects that researchers have focused on to identify their potential impact on RE-related activities, where the majority of the prior studies were based on academia. By conducting this survey, we wanted to identify the \textbf{industry} perspective on which of these human aspects software practitioners in the industry consider highly important when involved in RE-related activities.  Hence, we asked our survey participants to rate their agreement related to the statement, \emph{"The success of the requirements engineering greatly depends on the people involved in the requirements engineering activities as their performance varies from one another"} where 45\% of the participants strongly agreed, while 41.4\% of them agreed with the statement. 11.7\% mentioned it as somewhat agree while 0.9\% was neutral about the statement. Only 0.9\% mentioned that they somewhat disagree with the statement whereas no one has disagreed or strongly disagreed to the statement which indicates that the  majority believes \emph{"people"} as an important aspect in RE-related activities. 
\par As we wanted to know participants' perspective on the influence of human aspects on RE-related activities, we provided a list of human aspects that we identified from our previous study and asked to rate to what extent those human aspects are important considering its influence on RE-related activities based on their perspectives. As shown in figure \ref{Fig 1: Human aspects in RE}, from the given list of human aspects, participants rated that some are extremely and very important considering its influence when involved in RE-related activities. From the perspectives of large number of participants, \emph{motivation (90.9\%)} and \emph{domain knowledge (90.7\%)} were considered highly important aspects, and \emph{attitude (88.3\%)}, \emph{communication skills (82.9\%)}, and \emph{personality (56.8\%)} were other human aspects where a considerable amount of participants mentioned as very/extremely important. From the 111 participants, 100 of them had experience in working with geographically distributed teams -- among these, only 26\% mentioned that \emph{geographic distribution} of the individuals is an influencing aspect on performing RE-related activities. In the studies \cite{RN1606} and \cite{RN1607} that specifically focus on challenges in RE during global software development, they have mentioned that social and cultural aspects of RE need to be taken into consideration as there are a significant amount of risks and challenges specially in RE-related activities in global software development. \emph{Emotions (32.4\%)} and \emph{cultural diversity (18.9\%)} were also rated as moderately important human aspects, where the majority of the participants mentioned \emph{gender} and \emph{age} as what they felt were the least important human aspects considering its influence when carrying out RE-related activities, with percentages of 6.3 and 11.7 respectively.
\par Apart from the given list of human aspects derived from \cite{RN1600}, we wanted to know if there are any other human-related aspects that our survey participants think as important referring to its influence on RE-related activities. 33 participants mentioned several other human-related aspects that they consider as important when conducting RE-related activities. Four of them repeated the same aspects that we have already given in the list, such as domain knowledge, culture, listening (communication skill) and working time zone (geographic distribution), whereas the rest came up with other human-related aspects, including \emph{experience, flexibility/adaptability, mental state, collaboration/interaction, language, marital status/family, creativity, leadership skills, empathy, health, independent thinking, race \& religion, ethnic background, out of the box thinking, designing thinking mindset, ability to context/task switch, directness, availability, sobriety, pessimism and ability to learn}. The experience was mentioned by several (4 times), and flexibility/adaptability, collaboration/interaction, and language were mentioned 3 times by the participants. Marital status, creativity, leadership skills and design thinking mindset were mentioned twice, and the other aspects were mentioned by one participant. 
\par As shown in table \ref{TABLE 2: Other human-related aspects}, several participants explained why their identified human-related aspects are important and how they influence their performance when involved in RE-related activities in the open-text answers. For example, R54, a senior business analyst  mentioned, \emph{"marital status and kids should have included in the list as I believe it is quite impactful especially with the current work from home (WFH) situation"}. R73, an IT consultant explained the influence of having independent thinking as \emph{"independent thinking is, as in most software roles, a huge factor. The ability to think on their feet and do so effectively, without leaning on other members of the team is a key attribute to perform RE-related activities effectively"}. However, only 07 participants gave brief explanations, while others indicated the human aspects without providing any reasons related to its influence.
\begin{table}[]
\centering
\caption{\centering Influence of other human-related aspects}
\label{TABLE 2: Other human-related aspects}
\resizebox{0.9\columnwidth}{!}{%
\begin{tabular}{@{}llll@{}}
\toprule
\multicolumn{1}{l}{\textbf{} }      & \textbf{Job Title} & \textbf{Other Human-related Aspects} & \textbf{Influence}                                                \\ \midrule
\multirow{2}{*}{\begin{tabular}[c]{@{}l@{}}  R21\end{tabular}}  & \multirow{2}{*}{\begin{tabular}[c]{@{}l@{}}  Senior Business Analyst \end{tabular}} 
 & Flexibility/Adaptability & \begin{tabular}[c]{@{}l@{}}Helps to cope with the situations when specially working in \\ an agile environment  \end{tabular}\\ 
 & & Design thinking mindset & Somewhat important to come-up with better solutions\\\cmidrule(l){3-4}\\
R41 & Business Analyst & Experience & Helps to manage difficult customers \\\cmidrule(l){3-4}\\
\multirow{2}{*}{\begin{tabular}[c]{@{}l@{}}  R54 \end{tabular}}  & \multirow{2}{*}{\begin{tabular}[c]{@{}l@{}}  Senior Consultant \end{tabular}} 
 & Marital status/family  & Specially important with the current WFH situation  \\
 & & Ability to context/task switch & Helps to react and handle multiple tasks at the same time\\ \cmidrule(l){3-4}\\
 R72   & Business Analyst & Leadership skills & \begin{tabular}[c]{@{}l@{}} Without proper leadership, a team would not be able \\ to work effectively \end{tabular}\\\cmidrule(l){3-4}\\
R73 & IT Consultant & Independent thinking & \begin{tabular}[c]{@{}l@{}}A key attribute to perform RE-related activities effectively without\\ leaning on other team members \end{tabular}  \\\cmidrule(l){3-4}\\
R60 & IT Manager & Ability to learn & Helps to work effectively in new environments \\\cmidrule(l){3-4}\\
R88 & Software Engineer & Collaboration/interaction & Helps to increase a healthy competition among team members\\

\bottomrule
\end{tabular}%
}
\end{table}

\subsection{RQ2 -- Factors that  motivate individuals to perform their RE-related activities effectively} \label{section 4.3}
Our SLR found motivation to be an under-researched human aspect in terms of its impact on RE-related activities. Thus in our survey, we wanted to further investigate the impact of motivation on RE-related activities. There, we identified a set of factors that motivate software practitioners when they involved in RE-related activities. Moreover, using open-ended responses, many explained \emph{why} certain factors are important and \emph{how} they impact their motivation to perform RE-related activities. 
As discussed in section \ref{section 4.2}, we identified the importance of human aspects during RE, and among them, based on participants' perspective, motivation is rated as the highest influential human aspect when performing RE-related activities. By using STGT to analyse open-text answers to the question \emph{"In your opinion, what are the factors that motivate you to perform effectively in requirements engineering activities? Please explain briefly why?"} we identified a set of influential factors for individuals' motivation when involved in RE-related activities. 
We then categorized these motivation factors as collaborative \& human aspects and technical aspects. The factors directly related to the people involved in RE-related activities are categorized as "collaborative \& human aspects", and the factors related to technical outcomes of the RE-related activities are categorized as "technical aspects". \textcolor{black}{While we asked our survey participants which factors motivate them to perform effectively, higher motivation does not always necessary mean higher work performance.}

\begin{figure*}[b]
  \includegraphics[width=\linewidth]{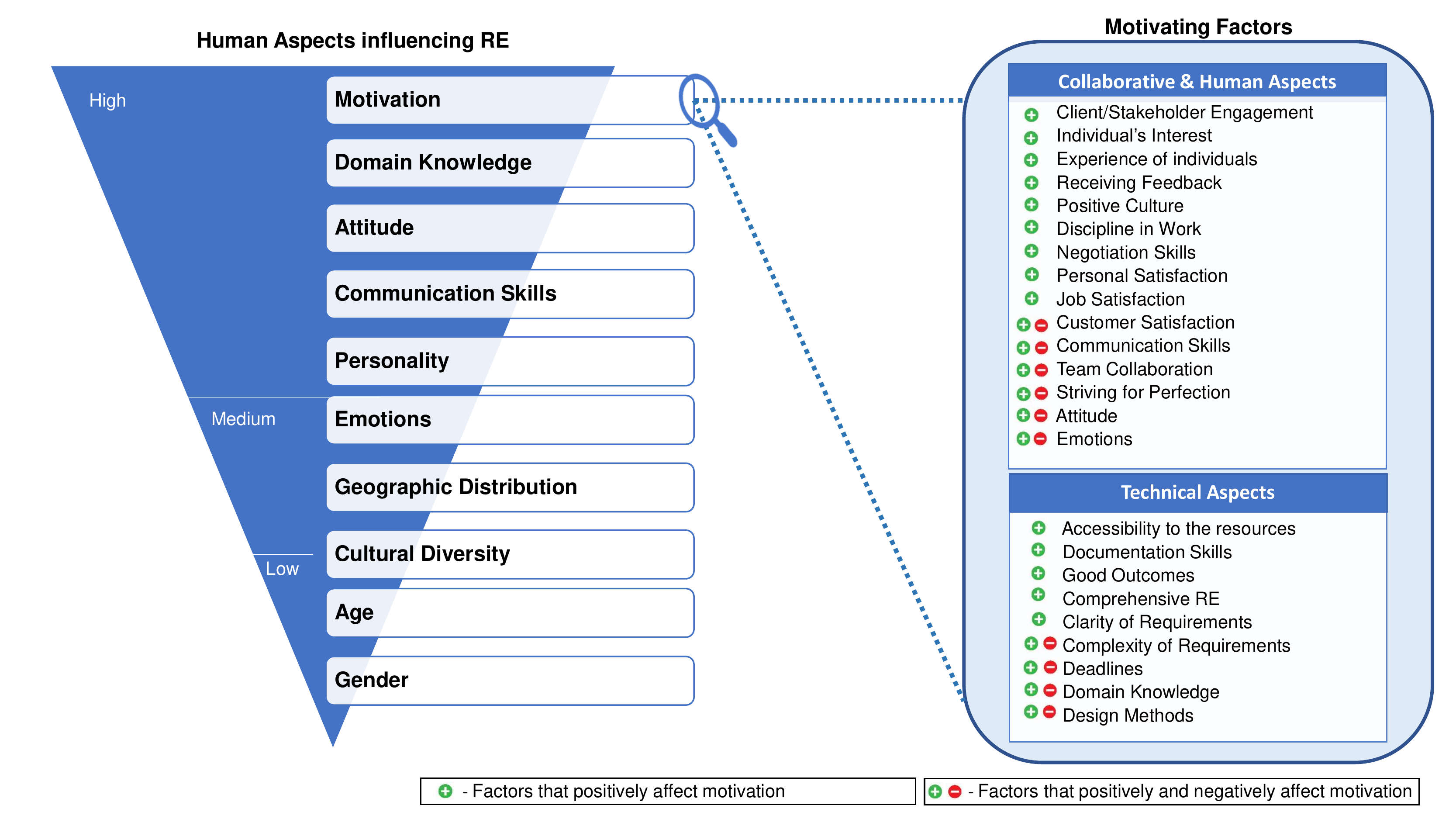}
  \caption{Human aspects influencing RE; identified through quantitative analysis (LHS), and in depth breakdown of motivation factors; identified through qualitative analysis (RHS)}
  \label{Fig 6: Importance of Human aspects}
\end{figure*}

As shown in figure \ref{Fig 6: Importance of Human aspects}, the left-hand side (LHS) of the diagram presents the human aspects influencing RE-related activities as rated by the participants (section \ref{section 4.2}). The right-hand side (RHS) shows the factors that affect individuals' motivation when involved in RE-related activities. While analysing these open-text answers, we identified that from practitioners' perspective, human aspects such as communication skills, emotions, attitudes, and domain knowledge can affect individuals' motivation when involved in RE-related activities. Hence, to show that these human aspects act as motivation factors, we included them in the both sides of the diagram. By conducting further analysis of these open-text answers, we were able to identify the impact of these motivation factors and according to that, the majority of the factors positively affect the motivation while some factors have both positive and negative affect on motivation when involved in RE-related activities. 

\par As shown in Table \ref{Table 5: impact of motivation factors}, the majority of our survey participants mentioned \textbf{Team collaboration} as their key motivation factor to perform RE-related activities. They also mentioned that better engagement within the team, proper response/support from the team, a friendly team environment, interaction with the team members and having diverse, empowered teams all help to motivate them to perform RE-related activities. Moreover, many have explained why it is important to have a good team collaboration. For example, R28, a business analysts mentioned, \emph{"When I get to collaboratively work with all required team members (example: doing a design sprint can easily solve a problem)"}, as the motivation factor and explained the importance of team collaboration in solving a problem. Apart from easy problem solving, participants have mentioned several other reasons such as, the ability to conduct RE-related activities better, and on-time product delivery in line with customer expectations by considering team collaboration as an important motivation factor. \textbf{Personal satisfaction} is another key motivation factor where participants mentioned that self-satisfied work, doing something useful, positive appreciation, recognition from the organisation, salary \& incentives, promotions and understanding their importance/worth,  make them more motivated to perform their RE-related activities.  \textbf{Customer/client/Stakeholder engagement} was also identified as another key motivation factor. Respondents  mentioned that good relationship \& close contact with clients, their frequent visits, commitments and collaborations make them motivated which will lead them to get correct, clear requirements more quickly. For example, R21, a senior business analysts explained how customer/stakeholder engagement motivates him/her as \emph{"Collaboration and positive culture  among overall stakeholders.-This gives us the motivation to do a proper job and be happy and satisfied within the process}. Furthermore, participants consider having good \textbf{communication skills} also motivate them to perform RE-related activities and stated that they they believe clear, friendly and direct communication with their team, clients or organisation can help solving problems more easily and increase their work efficiency. 

\begin{table*}[]
\caption{Impact of the motivation factors on performing RE-related activities}
\label{Table 5: impact of motivation factors}
\resizebox{\linewidth}{!}{%
\begin{tabular}{@{}clll@{}}
\toprule
\textbf{Category}                                                                   & \textbf{Motivation Factor}                                            & \textbf{Impact on RE/SE} & \textbf{Responses} \\ \midrule
\multirow{15}{*}{\begin{tabular}[c]{@{}l@{}} Collaborative \&\\ Human \\ aspects\end{tabular}}                                                      & Team collaboration                                                & \begin{tabular}[c]{@{}l@{}} Friendly collaboration will make the process fun and effective \\ Proper support \& opinion of the team will easily solve the problems by making decisions\\ and  solutions together Cooperation among team will increase the ability to work well \\by helping each other, result in final product in lined with customer expectations \\ Working with the team will be helpful in doing RE clearly, making the product delivery \\on-time \\ Working with like-minded individuals leads to less problems \\ Helps to come up with better solutions with the collaboration of diverse people \end{tabular}  &  25 \\  \cmidrule(l){2-4} 
    & Personal satisfaction                                            & \begin{tabular}[c]{@{}l@{}} Salary, incentives, promotions, recognition motivate individuals to achieve any goal \\ Self-satisfaction of doing something useful will make practitioners finish their\\ work correctly \\ Positive appreciation may helpful in further development \end{tabular} & 15 \\ \cmidrule(l){2-4}

    & \begin{tabular}[c]{@{}l@{}} Customer/Client/Stakeholder\\ engagement   \end{tabular}                                         & \begin{tabular}[c]{@{}l@{}} Good relationship with client, help in getting all the requirements clearly \\ Frequent visit of the clients increase the opportunities to involve with them, \\ understand them and advantage of gathering more requirements  \\ Collaboration with clients will be helpful in doing proper job, making the clients happy\\ \& satisfied \\ Proper customer engagement, do not get vague requirements which is important in \\ software development \end{tabular} & 11                                                                                                        \\ \cmidrule(l){2-4} 
        & Communication skills                               
        & \begin{tabular}[c]{@{}l@{}} Good communication skills increase the clarity with the clients \\ Increase the ability to communicate directly which will make the work efficient \\ Helps to work with clear mindset \\ Improve the ability to discuss the steps without any fear and obtain clear descriptions on \\ what the system suppose to do which is essential for good performance \\ Helpful in identifying the problems easily \end{tabular}  & 10                                               \\ \cmidrule(l){2-4} 
             & Individual's interest                                
             & \begin{tabular}[c]{@{}l@{}} Helps to achieve goals \& increase the quality of the service \\ Makes the practitioners create novel/ highly practical features \end{tabular}      & 10      \\\cmidrule(l){2-4} 
        & Customer satisfaction                                     
        & \begin{tabular}[c]{@{}l@{}} 100\% satisfied customers impact on the organisation \& its functions \\ Increase the practitioners' desire on developing features wanted by the customers\end{tabular}     & 7                        \\ \cmidrule(l){2-4} 
                & Experience of individuals                                                                                                                                           & \begin{tabular}[c]{@{}l@{}} Having more experience increase the quality of service which will be impact on the\\ final product \end{tabular}  & 6                                                  \\ \cmidrule(l){2-4} 
            & Job satisfaction                                                                                           & \begin{tabular}[c]{@{}l@{}} Leads to happiness \& satisfaction which motivates to do new activities \\ Complete the tasks successfully \end{tabular}  & 6                                                                                               \\ \cmidrule(l){2-4} 
         & Attitude                                                 
         & \begin{tabular}[c]{@{}l@{}}Attitude of the team/client/top management motivate to work collaboratively  \end{tabular}                                                                                   & 3     \\\cmidrule(l){2-4}                     & Receiving feedback                                                                                                               & \begin{tabular}[c]{@{}l@{}} Customer/team feedback  make people do something better \\ Increase customer/team engagement \end{tabular} & 3                                                          \\\cmidrule(l){2-4}
         & Positive work environment                                                                                                              & \begin{tabular}[c]{@{}l@{}} Make the practitioners do their job properly\\ Increase the satisfaction with the process \end{tabular}    & 3                                                             \\\cmidrule(l){2-4}
         
        & Negotiation skills                                          & \begin{tabular}[c]{@{}l@{}} Important to have between the team \& clients or negotiate among different approaches\\ to achieve specific goals \end{tabular}  & 2                                                  \\\cmidrule(l){2-4}
         & Striving for perfection                                                                                 & \begin{tabular}[c]{@{}l@{}} Increase the perfection in all tasks \\ The job gets done with the best ability and there will be no flaws in the tasks \end{tabular}    & 2           \\\cmidrule(l){2-4}

& Emotions                                         & \begin{tabular}[c]{@{}l@{}} Specially impact when deadlines are close and desperately want to finish tasks \end{tabular}  & 2 \\\cmidrule(l){2-4}

& Discipline in work                                        & \begin{tabular}[c]{@{}l@{}} Improve the ability to function without rely on others \\ Make the fellow practitioners relief and focus on their own work \end{tabular} & 1 

\\ \midrule
\multirow{9}{*}{\begin{tabular}[c]{@{}l@{}} Technical \\aspects\end{tabular}} & \begin{tabular}[c]{@{}l@{}} Domain knowledge\end{tabular}                                                                                                   & \begin{tabular}[c]{@{}l@{}} Having required domain knowledge leads to think about requirements in \\a different way \\ Make practitioners pay more attention to the details \\ Get the clear idea on what need to be implemented \end{tabular}  & 13                                                                \\\cmidrule(l){2-4}
     & Clarity of requirements                                                                                                                       & \begin{tabular}[c]{@{}l@{}} Helps to understand requirements and perform the analysis well \\ Massively reduce the pain, effort, disruption in the later phases of the software \\development cycle \\ Impact on the improvements of the features,  Helps to reveal true requirements \end{tabular}    & 9          \\\cmidrule(l){2-4}
    
     & Comprehensive requirements engineering                                                                   & \begin{tabular}[c]{@{}l@{}} Perform RE-related activities properly by considering all the requirements \\  Make the rest of the development cycle easier \\ Helps to deliver the product on-time \end{tabular}   & 8                                               \\\cmidrule(l){2-4} 
& Accessibility to the resources                                                                                       & \begin{tabular}[c]{@{}l@{}} Perform  RE-related activities effectively with the easy access to the information \\ Having good prototype tools increase the ability to create designs quickly \\ Make practitioners efforts work and worth \end{tabular}      & 6                                             \\\cmidrule(l){2-4}
 & Good outcomes                                                           & \begin{tabular}[c]{@{}l@{}}Good end results make the clients satisfied \\ Improve the determination to see things workout at the end\end{tabular}        & 4                                                                                    \\\cmidrule(l){2-4}
& Complexity of the requirements/scope                                                  & \begin{tabular}[c]{@{}l@{}} Make it more interesting to the software practitioners \\ Improve their sense of satisfaction \end{tabular}      & 2       \\\cmidrule(l){2-4}
        & Deadlines                                                                                          & \begin{tabular}[c]{@{}l@{}} Make practitioners sign-off from the tasks, which will generate revenue   \end{tabular}  & 2                                                           \\\cmidrule(l){2-4}
          & Documentation skills                                                                      & \begin{tabular}[c]{@{}l@{}} Make the work easier by writing quality requirements \\ Helps to avoid issues with team members \end{tabular}     & 2                    \\\cmidrule(l){2-4}
 & Design methods                                                              & \begin{tabular}[c]{@{}l@{}} Improve the understandability of the requirements via more visual than descriptive \end{tabular}    & 1                                                     
     \\ \cmidrule(l){1-4} 
\end{tabular}%
}
\end{table*}

 \par \textbf{Individual's interest} towards the project, including its domain, requirements, novel features, interest in working with the team,  and interest in what they do, make software practitioners more motivated. As a result, this will increase the quality of their RE-related activities they are involved. Apart from these, \textbf{customer satisfaction}, the \textbf{experience of individuals}, \textbf{job satisfaction}, the \textbf{attitude} of the team, client or top management, and \textbf{receiving feedback} from the clients or the team were other commonly mentioned motivation factors by our survey participants. For example, R68, a senior specialist mentioned \emph{"successfully implementing correctly prioritized requirements in a release results in happy stakeholders and a better functioning organisation overall"} as his/her main motivation while involved in RE-related activities. Considering the technical aspects, \textbf{domain knowledge} was mentioned as a key motivation factor by the majority of respondents. They mentioned that having the required knowledge for the project, competency in project subject motivates them to think about requirements differently, and gives them time to pay more attention to the details of the project.  \textbf{Clarity of the requirements} is another motivation factor where this was explained as clear, well-thought-out requirements motivating them to perform RE-related activities, resulting in better requirements analysis, improvement of the features, and massively reducing the pain, efforts or disruption later phases of the software development cycle. \textbf{Comprehensive requirements engineering} is identified as a motivation factor where participants  mentioned that performing RE-related activities properly motivates them to consider all the requirements making the rest of the software development tasks easier, and helps to deliver the product on time. \textbf{Accessibility to the resources} such as tools, information or new technology also helps to motivate many respondents to perform RE-related activities. For example, it was mentioned that having good prototyping tools leads them to create their candidate designs more quickly for feedback, resulting in more effective RE performance. \textbf{Good outcomes}, including smooth project executions and clearness of the solutions, act as a motivation factor as they lead to greater customer satisfaction. The \textbf{Complexity of the requirements/project scope} also motivates many surveyed practitioners as it results in increasing their interest and satisfaction in their projects' RE-related activities. \textbf{Meeting deadlines} of the projects motivates some practitioners as these will generate the revenues for the project, where some get motivated with writing quality requirements (\textbf{documentation skills}) and mentioned that it makes their work easier. 
\par Apart from the above mentioned motivation factors, there are some additional motivation factors mentioned by a few participants. Positive culture, emotions, negotiation skills, strive for perfection, and discipline in work was mentioned by one or two participants that have been categorized under collaborative \& human aspects. We had to remove some open-text answers due to less/no explanations, where participants have only mentioned one or two words such as "morale, management support, professional involvement, good leads and etc.  Considering all the identified motivation factors, we identified that the majority of the participants (59.5\%) are  motivated by collaborative \& human aspects whereas, 40.5\% of participants get motivated by technical aspects. The participants explained how these reasons would impact on performing RE-related activities, and the success of the overall project. Table \ref{Table 5: impact of motivation factors} shows a summary of the identified motivation factors and their impact on RE and SE that will directly or indirectly affect the outcome of the overall project. With these findings, it is shown that from software practitioners' perspective, there are number of collaborative \& human-related aspects that impact individuals motivation when involved in RE-related activities and further investigations are needed for most of these aspects. 

\subsection{RQ3 -- What personality characteristics are important for effective performance of RE-related activities} \label{section 4.4}
\begin{figure}[]
   \includegraphics[width=0.7\linewidth]{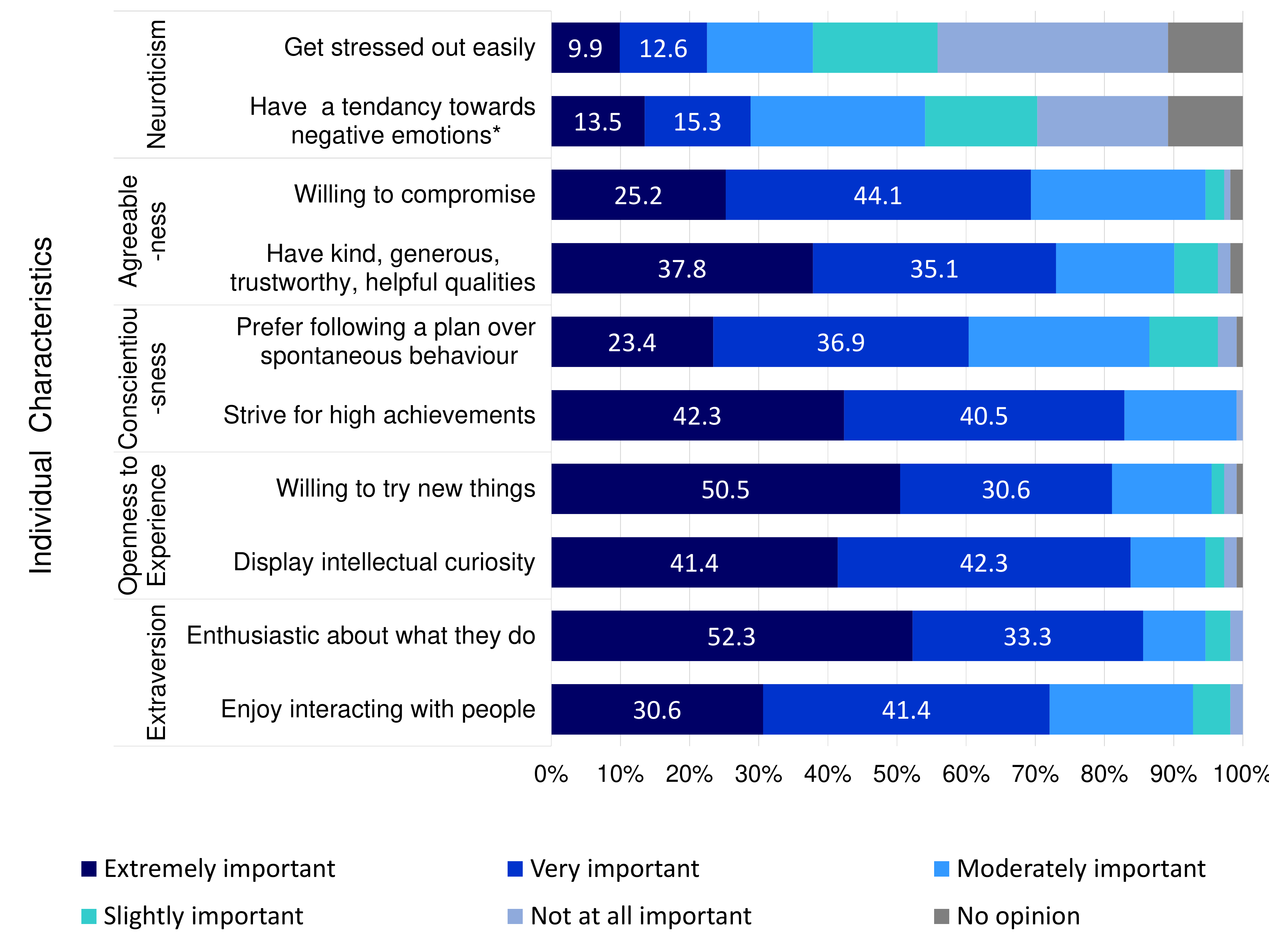}
  \caption{Individual characteristics that are important when involved in RE-related activities }
  \label{Figure 05: Individual characteristics}
\end{figure}

Our SLR indicated motivation and personality aspects are under-researched human aspects in terms of their impact on RE-related activities. Thus we wanted to get software practitioners' perspective on the importance of personality characteristics on RE-related activities. There, we identified that the majority of participants think that the success of RE-related activities depends on the people involved in those activities as their different characteristics impact RE-related activities in various ways.  We asked our survey participants to rate their agreement with the statement, \emph{"Differences in characteristics, behaviours, personal habits, skills of the people involved in the requirements engineering activities affect the requirements engineering process"}, 31.5\%  of participants strongly agreed, whereas 36.9\%  agreed to it. 27\% rated it as somewhat agree whereas 4.5\% of participants were neutral about it. No one disagreed with the statement as there was 0\% for somewhat disagree, disagree and strongly disagree. This indicates that our survey respondents believe that the success of RE-related activities greatly depends on the people involved in those RE-related activities and their characteristics considerably affect  its success. This statement is related with one of the most commonly used definitions of personality \cite{RN2976} \cite{RN2975}, and indirectly this indicates that software practitioners agree that the personality of individuals involved in RE-related activities has an impact on the success of it. To identify what are the key individual human characteristics that are important to the people involved in RE-related activities, we used a set of individual characteristics from the well-known "Five Factor Model" of personality, representing five broad personality dimensions that describe a range of characteristics of individuals. We have considered ten characteristics and asked the participants to refer to themselves and rate the importance of the characteristics when conducting RE-related activities. 
As shown in figure \ref{Figure 05: Individual characteristics}, among the characteristics, \textbf{enthusiasm about what they do} is the one that is rated as of the highest importance, with 52.3\% as extremely important and 33.3\% as very important. \par The second highest important characteristic is \textbf{display intellectual curiosity}, with a rating of 41.4\% as extremely important and 42.3\% as very important. \textbf{Strive for high achievements} is another important characteristic with  percentages of 42.3 (extremely important) and 40.5 (very important). Next in line is \textbf{willing to try new things} as 50.5\% of participants have mentioned it as extremely important whereas 30.6\% of them mentioned it as very important. \textbf{Having kind, generous, trustworthy, helpful qualities} is also rated as extremely important by 37.8\% and very important by 35.1\%. \textbf{Enjoy interacting with people} is also an important characteristic when performing RE-related activities, where 30.6\% rated it as extremely important and 41.1\% as very important. Characteristics such as \textbf{willing to compromise} and \textbf{prefer following a plan over spontaneous behaviour} tend to be moderately important, with the ratings of 25.2\% and 23.4\% as extremely important, respectively. 

According to our survey participants, characteristics such as \textbf{have a tendency towards negative emotions} and \textbf{get stressed out easily} do not have much importance when they are involved in RE-related activities. 33.3\% of participants have rated get stressed out easily as not at all important while 18.9\%  rated having a tendency towards negative emotions as not at all important. All the other above mentioned characteristics were rated with very low percentages as not at all important. For example, only 0.9\% rated strive for high achievements as not at all important whereas enthusiasm about what they do, willing to try new things, enjoy interacting with people were mentioned as not at all important by only 1.8\%, which indicates that majority of the individual characteristics are highly important when involved in RE-related activities. 

We also wanted to identify what are the important human characteristics that software professionals expect to see from their team members when involved in RE-related activities. Hence, we used the same set of characteristics and asked them to rate each considering its importance to have in their team members. The results are quite similar to the importance of their individual human characteristics described above. However, some software practitioners seem to believe that all the given characteristics are highly important to have in their team members compared to having only some themselves. 
\begin{figure}[]
  \includegraphics[width=0.7\linewidth]{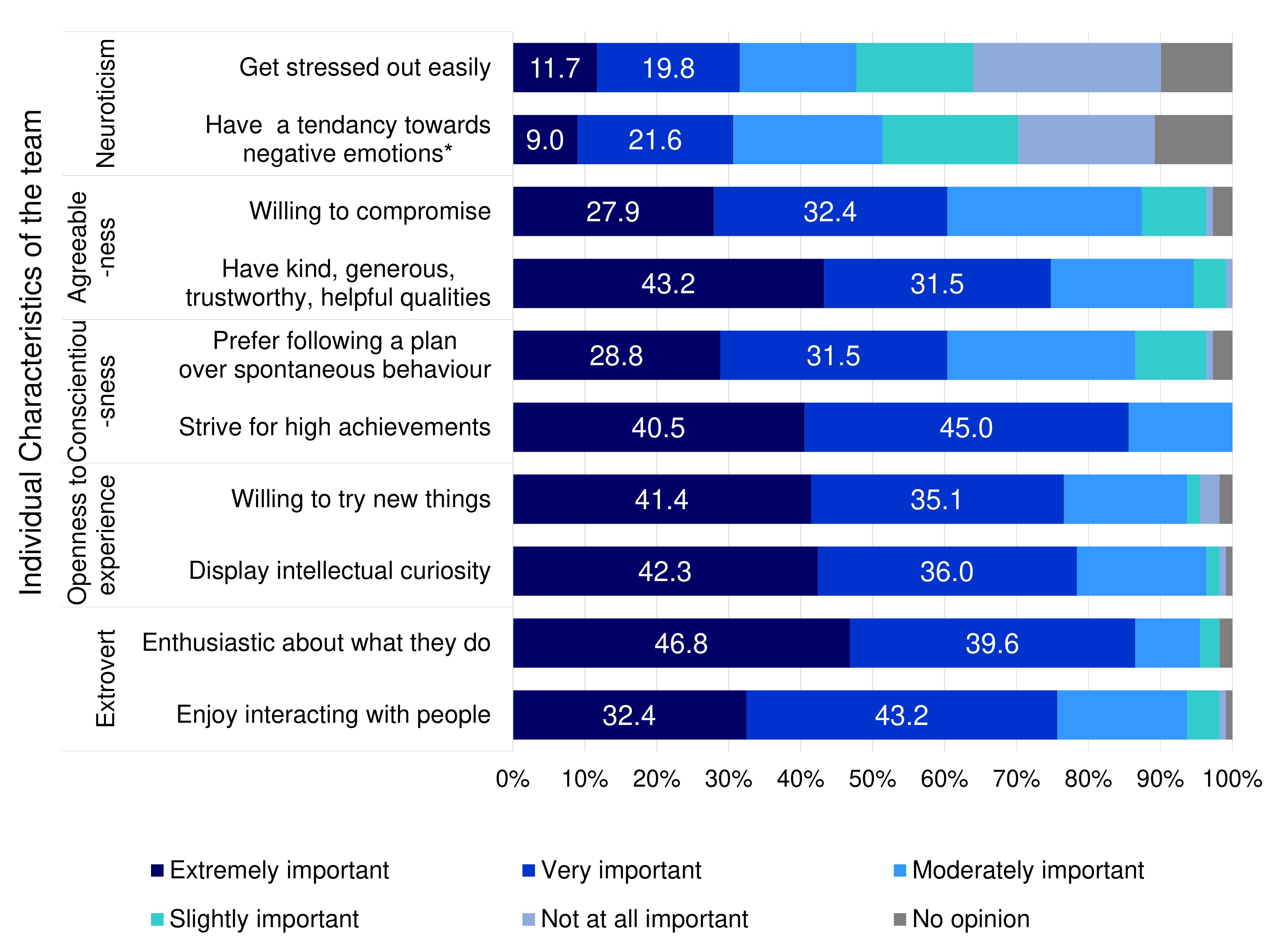}
  \caption{Characteristics that are important in team members when involved in RE-related activities}
  \label{Fig 5: Team characteristics}
\end{figure}

\par As shown in figure \ref{Fig 5: Team characteristics}, 86.5\% of the participants mentioned, that the \textbf{enthusiasm about what they do} is extremely or very important whereas 85.6\% mentioned \textbf{strive for high achievements} is extremely or very important to have in their team members. \textbf{Display intellectual curiosity}, and \textbf{willing to try new things} were also rated as extremely or very important characteristics with percentages of 78.4 and 76.6 respectively. \textbf{Enjoy interacting with people} and \textbf{have kind, generous, trustworthy, helpful qualities} were then rated as extremely or very important characteristics with 75.7 and 74.8 percentages respectively. Participants consider \textbf{prefer following a plan over spontaneous behaviour} and \textbf{willing to compromise} are equally important characteristics as 60.4\% of participants rated both as extremely or very important. Considering the last two characteristics, 31.5\% mentioned the \textbf{get stressed out easily} is extremely or very important whereas, 30.6\% of the participants mentioned \textbf{have a tendency towards negative emotions} is extremely or very important. This indicates that according to the participants' perspective, having majority of these personality characteristics within their team members are more important than having these characteristics in themselves. Apart from the above-mentioned characteristics, we found some additional characteristics that the participants have mentioned as important to have in people involved in RE-related activities via their open-ended survey responses. The majority  mentioned that the listed human characteristics are enough, but some  suggested a few other characteristics such as \emph{being confident}, \emph{attentiveness}, \emph{patience}, \emph{good empathy},
\emph{cultural sensitivity}, \emph{tardiness}, \emph{good listening skills} and \emph{adaptability} as important characteristics to have within the people involved in RE-related activities.

\begin{table*}[]
%\centering
\caption{Factors that make individuals less effective when involved in RE-related activities and their impact}
\label{TABLE 6: less effectiveness factors}
\resizebox{\linewidth}{!}{%
\begin{tabular}{@{}clll@{}}
\toprule
%\footnotesize
\textbf{Category}          & \textbf{Factor}                                         & \textbf{Impact on RE/SE} & \textbf{Responses}                                \\ \midrule
\multirow{14}{*}{\begin{tabular}[c]{@{}l@{}}Collaborative \&\\ Human\\ aspects\end{tabular} }   & Communication issues                                                            & \begin{tabular}[c]{@{}l@{}} Lack of communication/ miscommunication of the team/clients \\ impacts on gathering correct requirements \\ Do not get clear requirements \\ Have to spend more time and effort to capture right information \end{tabular}  & 23  \\
 
 & Management issues                                                      &  \begin{tabular}[c]{@{}l@{}} Organisational politics impact on product definition \& decision making \\ Not giving recognition for hard work makes people demotivated \\ Wrong assumptions of the management makes the projects complicated \\ Lack of encouragement provided results in less interest towards the project \end{tabular}   &     18                                                      \\ 
       
        & \begin{tabular}[c]{@{}l@{}}Nature of the clients/\\stakeholders/customers \end{tabular}                                                & \begin{tabular}[c]{@{}l@{}} Difficult to handle clients/ inexperienced clients make \\ RE-related tasks slow and time consuming \\ Lack of commitment of the clients makes the requirements \\ elicitation hard, result in negative outcomes \\\end{tabular}  & 16 \\
        & Team behaviour                                        &\begin{tabular}[c]{@{}l@{}} No unity among team/ no cooperation result in inability to carry out RE properly \\ Make the software practitioners not to work at all \\ Leads to impact on effective communication, productivity and negative outcome\end{tabular}     & 13                                                    \\

         & Heavy workload 
         & \begin{tabular}[c]{@{}l@{}} Prevent from concentrating on work properly \\ Increase less interest towards the tasks \end{tabular}  & 3 \\
         & \begin{tabular}[c]{@{}l@{}}Interaction with demotivated \\people \end{tabular} 
         & \begin{tabular}[c]{@{}l@{}} Demotivated people do not give their best to the process \\ Difficult to get their commitment \\ Increase the issues in the progress  \end{tabular}  & 3 \\
         & Negative attitudes 
         & Negative attitude of the people around prevent from working & 3 \\
         & Over desire to satisfy customer 
         & \begin{tabular}[c]{@{}l@{}} Make reluctant to enhance the project \& difficult to shift from one sprint to next \\ Doing more than required to solve the problem can put the actual \\ requirements at a risk \\ Create a lot more work   \end{tabular} & 3 \\
         
         & Individuals' nature
         & \begin{tabular}[c]{@{}l@{}} Being perfectionist sometimes leads to less effectiveness \\ Can impact on completing tasks quickly and releasing the products \\ Various personalities of the individuals may leads to difficulties in working \\ Conservative nature may impact on taking risks, trying new things and making \\ their ideas stand out   \end{tabular} & 3 \\
          & Personal issues 
         & \begin{tabular}[c]{@{}l@{}} Personal issues such as health, problems in family members make the distractions \\ from the work  \end{tabular} & 2 \\
         
          & \begin{tabular}[c]{@{}l@{}}Distracting working \\environment \end{tabular}
          & \begin{tabular}[c]{@{}l@{}} Prevent on having proper concentration on work \\ Decrease the interest towards work \end{tabular} & 2  \\

                               \midrule

\multirow{7}{*}{\begin{tabular}[c]{@{}l@{}}Technical \\aspects\end{tabular} }   
& Unclear requirements                                                         & \begin{tabular}[c]{@{}l@{}} When requirements are listed without specifying the actual need, it will be \\ difficult for the practitioners to derive the requirements \\ Unclear requirements can break some core logic of the design \\  Make the requirements gathering process hard and less effective \\ May not contribute to any revenue generation \end{tabular} & 18   \\

& Less domain knowledge            
    & \begin{tabular}[c]{@{}l@{}} Make RE-related tasks slow and impact on the features of the products \\ Can be a hindrance to the individuals career as well as fulfilling \\ customer demands \\Increase the inability to explain the necessity and needs of the requirements   \end{tabular} & 14                                          \\

     & Lack of resources                                           
     & \begin{tabular}[c]{@{}l@{}} Not having proper tools, required information or new technologies,  for RE \\activities can be slow and less effective \\ Tasks may remain unfinished  \end{tabular} & 9                                                            \\
     & Constant requirement changes                                              
     & \begin{tabular}[c]{@{}l@{}} Time-to-time requirements changes by the clients make the process less effective \\ Sudden and unreasonable requirements make the project harder to implement \\ May increase the problems in project budget\end{tabular}  & 9                                                         \\
    & \begin{tabular}[c]{@{}l@{}} Issues in the design \& \\the method \end{tabular}
    & \begin{tabular}[c]{@{}l@{}} Unnecessary charts, poor user stories will make the process less effective \\ Not following proper RE process may impact on the project outcome \\ Not using the most relevant software development methods may create difficulties \\ (e.g., using waterfall method for an agile project) \end{tabular} & 8                                                               \\
    & Strict deadlines 
    & \begin{tabular}[c]{@{}l@{}} May make the software practitioners' job tougher \\ Impact on the quality of the product \end{tabular} & 3                                                                \\ 
    & Unimportant Requirements & \begin{tabular}[c]{@{}l@{}} When there is no revenue generation/ value of the feature is not seen, makes\\ the individuals ineffective when involved in RE \end{tabular} & 2   \\
 \bottomrule
\end{tabular}%
}
\end{table*}

\subsection{RQ4 -- What factors make individuals less effective when performing RE-related activities} \label{section 4.5}
Effectiveness is known as the power to produce desired results and it is often measured as the quality of the desired results \cite{article}.  Though there are studies related to the effectiveness of the teams and organisations in SE, few studies are available on individual's effectiveness in the area of RE-related activities. We thus also wanted to identify what factors do practitioners think make individuals less effective when involved in RE-related activities. By analysing open-text answers to the question \emph{"In your opinion, what are the factors that make you ineffective when involved in requirements engineering? Please explain briefly why"} we identified a set of factors that make them less effective when involved in RE-related activities. Interestingly, most of these factors were just opposite factors that we identified as motivation factors and we were able to categorize these factors into  two similar groups -- collaborative \& human aspects and technical aspects while identifying that the majority of these factors related to the collaborative \& human aspect category. Our findings are shown in Table \ref{TABLE 6: less effectiveness factors}, and we discuss the most common responses for each category below.

\textbf{Communication issues} are the most reported factor (23 responses), where participants have mentioned that lack of effective communication or miscommunication in the team or with clients can be a great hindrance, making them less effective when involved in RE-related activities.  For example, R110, a senior analyst mentioned, \emph{"There are some customers/users, that are not effective in communication and cannot describe their needs. This type of person requires more time and effort to capture the right information"}. They explained that it would highly impact their requirements gathering work as they won't be able to get clear requirements from the clients. The \textbf{Nature of the clients/stakeholders/customers} is another factor that makes the practitioners less effective. Our survey respondents mentioned that clients who are difficult to handle, inexperienced clients, uncommitted clients, confused stakeholders or negativity in customers could impact their effectiveness negatively. For example, R20, an associate tech lead, mentioned, \emph{"When the customer has no idea about what he/she really wants"}, make them less effective. Moreover, it was reported that when there is a lack of commitment from  clients, it is hard to elicit their requirements. According to participants, the right handling of clients is very important for effective RE task performance as well as for the project. It was explained that having clients who constantly scrutinize the team is unhelpful, but it will be worse to have them completely out of the project, as the earlier their feedback is received, the less costly to incorporate it into the project. 

\textbf{Team behaviour} was identified as another factor that make individuals less effective. It was mentioned that when there is no unity among the team, no cooperation or irresponsible people in the team, these can lead to less effectiveness when involved in RE-related activities. For example, R77, a software engineer, mentioned, \emph{"Sometimes, I feel unproductive if I have to work with people that don't put as much effort into what they do like me"}. Moreover, participants reported that having the above team behaviours may impact negatively on the effectiveness of communication and productivity, and thus lead to negative outcomes of RE-related activities. 

Apart from these collaborative \& human aspects, there are some factors that we categorized as technical aspects that participants have mentioned, that they said make them less effective when involved in RE-related activities. Among them, \textbf{unclear requirements} and \textbf{less domain knowledge} are the highest reported factors. Participants  explained that when clients try to specify their requirements and list them without knowing actual system needs, it will be difficult to derive the actual software requirements which make them less effective. Considering domain knowledge, it was mentioned that when practitioners do not have sufficient knowledge of the domain, it will make RE-related activities slower and impact on features of the product. \textbf{Lack of resources} and \textbf{constant requirement changes} are another two factors that negatively impact on participants' effectiveness when involved in RE-related activities. When there are not enough resources, such as people, proper tools, information or use of appropriate technologies, it may make RE-related activities slow and less effective. Moreover, constant changes of the requirements such as late or major requirements changes by the clients, sudden, unreasonable changes of the requirements, or changes that come in the middle of major implementation work, will impact on the effectiveness of RE-related activities and create problems with the project budget and outcomes.

\subsection{RQ5 -- Participants' perspective on measuring the performance of individuals when involved in RE-related activities} \label{section 4.6}

Our previous RQs tried to identify software practitioners' perspectives on which human aspects most impact RE-related activities, both positively and negatively. This led us to wanting to get a better idea of how practitioners RE-related activity performance is actually measured, and how they think an individuals' performance should be measured when they perform RE-related activities. 

\begin{figure}[t]
  \includegraphics[width=0.7\linewidth]{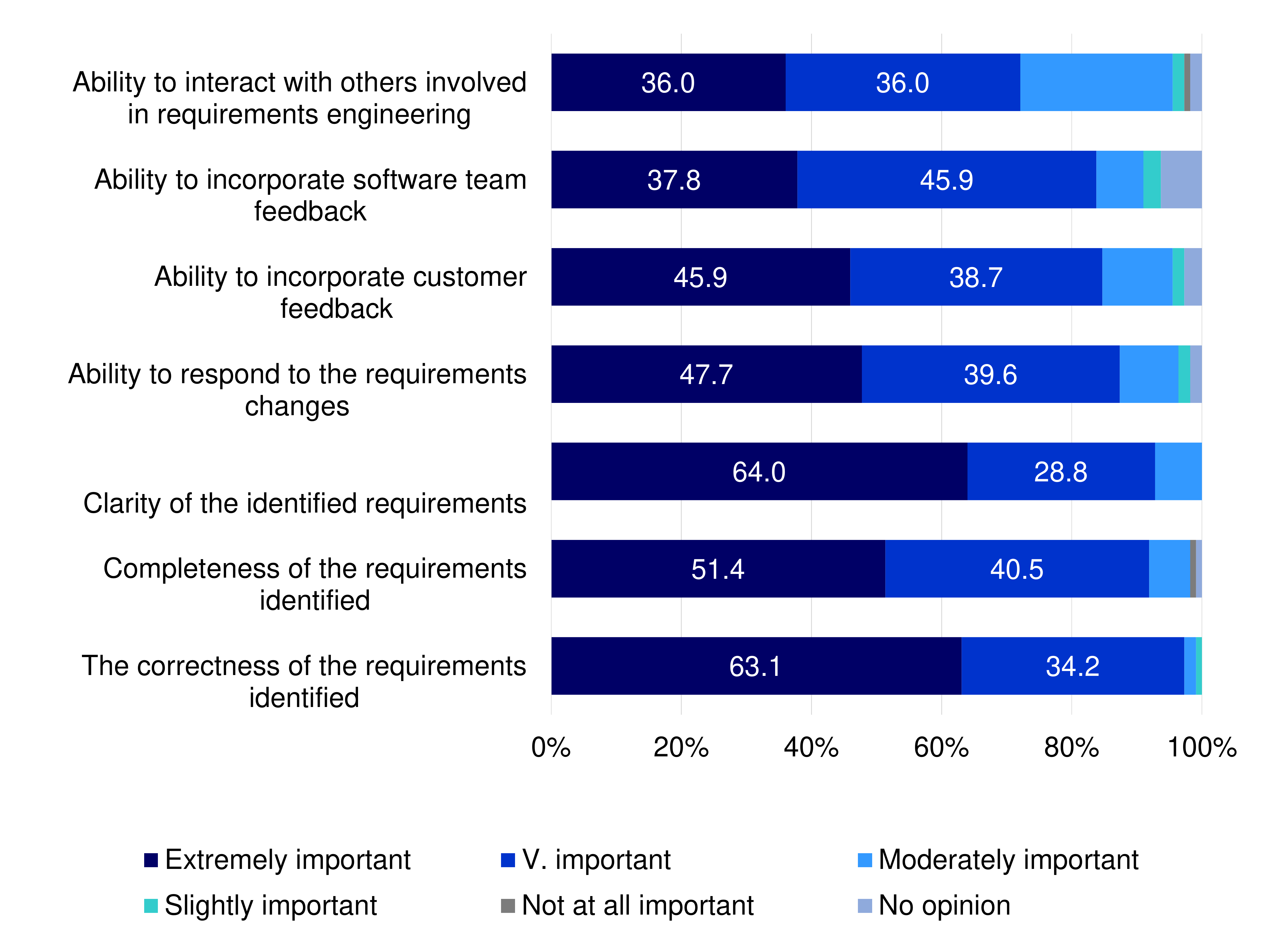}
  \caption{Factors important in measuring individuals' performance in RE-related activities}
 \label{Fig 3: Factors measuring performance in RE}
\end{figure}

As shown in figure \ref{Fig 3: Factors measuring performance in RE}, the majority (87.1\%) of our survey participants agree that all these factors are extremely/very important in \emph{measuring} performance of the people involved in RE-related activities. \textcolor{black}{Many of these are technical aspects of the work,} among them, the \emph{correctness of the requirements identified (97.3\%)}, the \emph{clarity of the identified requirements (92.8\%)}, the \emph{completeness of the requirements identified (91.9\%)}, and \emph{ability to respond to requirements changes (87.4\%)} were  mentioned as the most important factors when measuring an individuals' RE-related work performance. \textcolor{black}{Others are related to  human aspects related to the RE work}, including \emph{ability to incorporate customer feedback (84.7\%), ability to incorporate software team feedback (83.8\%)}, and \emph{ability to interact with others involved in RE (72.1\%)}, all rated as important factors for measuring performance when involved in RE-related activities. Apart from the given list of factors, we wanted to know if there are any other factors that  software practitioners think as important when measuring the performance of the individuals involved in RE-related activities. Hence, we included a follow-up open-ended question asking, \emph{"Are there any other factor(s) apart from the above list that you use to measure individual performance in requirements engineering activities, please mention:"}, where the majority mentioned "None". Only 3 participants mentioned other factors such as, \emph{being able to validate requirements/designs, ability to understand true requirements from users, and a good knowledge of the business (system) as a whole}.

\section{Discussion} \label{section 5}
\subsection{Key findings} \label{5.1}
\begin{table}[t]
\centering
\caption{\centering Key Findings (KF) of the study}
\label{TABLE 7: key findings}
\resizebox{0.8\linewidth}{!}{%
\begin{tabular}{@{}cll@{}}
\toprule
\footnotesize
\textbf{}  & \multicolumn{1}{c}{\textbf{Key Findings (KF)}}                                         & \multicolumn{1}{c}{\textbf{Section}}                                \\ \midrule
\multirow{1}{*}{KF1}   & \begin{tabular}[c]{@{}l@{}} Software practitioners believe that human aspects influence the performance of individuals \\when involved  in RE-related activities \end{tabular} 

& \ref{section 4.2}    \\
                                \midrule

\multirow{1}{*}{KF2}      & \begin{tabular}[c]{@{}l@{}} Based on participants' perspective, motivation, domain knowledge, attitude, communication \\skills and personality were  identified as highly important human aspects considering its influence\\ on their performance while emotions, cultural diversity, geographic distribution as moderately \\important aspects. Age and gender were mentioned as least important aspects which indicates\\ these two aspects make least influence on their performance \end{tabular}                                                       & \ref{section 4.2}                                                                                            \\ \midrule
\multirow{1}{*}{KF3}  & \begin{tabular}[c]{@{}l@{}} A set of factors were identified that motivate individuals when performing RE-related \\activities and the majority of them were belongs to the "collaborative \& human aspect"\\ category \end{tabular}        & \ref{section 4.3}         \\ \midrule

\multirow{1}{*}{KF4}  & \begin{tabular}[c]{@{}l@{}} Software practitioners get highly motivated by team collaboration, personal satisfaction, \\customer/stakeholder engagement, communication skills, individual's interest \\ when involved in RE-related activities.   \end{tabular}                                                         & \ref{section 4.3}                                                            \\ \midrule
\multirow{1}{*}{KF5}  & \begin{tabular}[c]{@{}l@{}} Domain knowledge, clarity of requirements, comprehensive RE,\\ accessibility to the resources, good outcomes, complexity of requirements, deadlines,\\ documentation skills \& design methods are also motivate software practitioners \\when involved in RE-related activities. \end{tabular}                                           &       \ref{section 4.3}                                                        \\ \midrule

\multirow{1}{*}{KF6}  & \begin{tabular}[c]{@{}l@{}} According to the participants' perspective, these motivation factors impact on performing RE \\activities and increase the quality of final product or service provided \end{tabular}                                                         &   \ref{section 4.3}                                                           \\ \midrule

\multirow{1}{*}{KF7}  & \begin{tabular}[c]{@{}l@{}} Majority of the software practitioners believe that RE-related activities greatly depends on the \\ people involved in the process and the differences in their personality characteristics, \\ impact on RE-related activities \end{tabular}                                                         &    \ref{section 4.4}                          \\ \midrule
\multirow{1}{*}{KF8}  & \begin{tabular}[c]{@{}l@{}} Software practitioners think that their individual personality characteristics such as\\ enthusiasm, intellectual curiosity, strive for high achievements, willingness to try new things,\\ enjoying interacting with people, and having qualities like kindness, trustworthiness, \\and generosity are highly important when involved in RE-related activities.  \end{tabular}                                                  &  \ref{section 4.4}                                                               \\\midrule
\multirow{1}{*}{KF9}  & \begin{tabular}[c]{@{}l@{}} Majority of the software practitioners also think that personality characteristics such as having \\tendency  towards negative emotions or get stressed out easily are not that much important\\ when involved in RE-related activities. \end{tabular}                                                         &   \ref{section 4.4}    \\ \midrule

\multirow{1}{*}{KF10}  & \begin{tabular}[c]{@{}l@{}} Software practitioners also consider that these personality characteristics are highly important \\to have in their team members.   \end{tabular}                                                         &   \ref{section 4.4}   
                                                                                             \\\midrule
 \multirow{1}{*}{KF11}  & \begin{tabular}[c]{@{}l@{}} There are factors that make the individuals less effective when involved in RE-related\\ activities and some of them were just opposite of motivation factors that we have identified.  \end{tabular}                                                         &      \ref{section 4.5}                                                        \\ \midrule

\multirow{1}{*}{KF12}  & \begin{tabular}[c]{@{}l@{}} Apart from the opposites of the motivation factors, nature of the clients/stakeholders, team\\ behaviour, constant requirements changes, heavy workload, management \& personal \\issues make the software practitioners less effective \end{tabular}                                                         &      \ref{section 4.5}                                                        \\ \midrule
\multirow{1}{*}{KF13}    & \begin{tabular}[c]{@{}l@{}} Considering the participants' perspective on measuring the performance of the individuals\\ involved in RE-related activities, the correctness, completeness and clarity of requirements need\\to be considered the most.  \end{tabular}                                                         & \ref{section 4.6}
                                                                                            \\ 
 \bottomrule
\end{tabular}}
\end{table}

Table \ref{TABLE 7: key findings} summarises the key findings from our survey relevant to each section. From the SLR we conducted earlier \cite{RN1600}, we identified a need for industry-focused studies to know software practitioners' perspectives on the influence of human aspects in RE-related activities. Hence, with this study, we have identified a set of human aspects that are important to consider when involved in RE-related activities, according to software practitioners. As \emph{motivation} and \emph{personality} have been well studied in SE in general but not as well investigated in RE in particular \cite{RN1600}, we will focus on these two human aspects in our future studies. Revealing that our current research focus aligns with the perspectives of software practitioners in the industry, motivation and personality were identified as highly important human aspects when involved in RE-related activities. We have identified that software practitioners get motivated by various factors when they are involved in RE-related activities.  We categorized these factors into two types, named \textit{collaborative \& human aspects} and \textit{technical} aspects where most of the motivation factors reported by our survey respondents are collaborative \& human aspects (Table \ref{TABLE 7: key findings} - KF3, KF4, KF5). For example, \emph{team collaboration} was recognised as the factor that motivates individuals the most. They said that better interactions within the team lead to better problem solving, and high-quality product delivery.

\par All these factors may impact on RE-related activities as well as overall software development outcomes. While these factors motivate individuals to perform RE-related activities, we identified that there are some factors  that make individuals less effective when performing RE-related activities.Some of these tend to be the opposites of motivation factors where \emph{lack of proper communication} is identified as the major factor that makes individuals less effective. Apart from that, we identified various factors that make the software practitioners less effective when involved in RE-related activities (Table \ref{TABLE 6: less effectiveness factors}). These were claimed to make RE-related activities slower, less accurate and less productive. Based on these findings, we identified that various human aspects appear to have the ability to reduce an individual's motivation or less effective when involved in RE-related activities. Moreover, we wanted to identify industry perspectives on personality and a set of questions were focused on individual and team personality characteristics.  According to our surveyed software practitioners, the individuals involved in RE-related activities should be highly enthusiastic about what they do, display intellectual curiosity, strive for high achievements, be willing to try new things, enjoy interacting with people, and have kind, generous, trustworthy, helpful qualities. The practitioners consider that having these characteristics are highly important for themselves as well as for their teams. Among the list of characteristics, willingness to compromise and preferring to follow a plan over spontaneous behaviour was said to be moderately important.  Having a tendency towards negative emotions and getting stressed out easily are considered the least important characteristics. However, when comparing themselves and their teams, software practitioners tend to consider that having all these characteristics are important in their team-mates if not all themselves.

\subsection{Comparison to Related Work} \label{5.2}
\begin{table*}[]
\caption{Comparison of the study with the existing related work}
\label{Table 5: comparison with related work}
\resizebox{\linewidth}{!}{%
\begin{tabular}{@{}cllll@{}}
\toprule
\textbf{Study}          
& \textbf{Human Aspect(s)}                                            
& \textbf{Context of Study} 
& \textbf{Method/Approach} & \textbf{Major outcomes} \\ \midrule

Coughlan and Macredie \cite{RN2952} 
& Communication & \begin{tabular}[c]{@{}l@{}} Global software \\development (GSD) \\-requirements elicitation \end{tabular} 
& \begin{tabular}[c]{@{}l@{}} Academic-based \\-a methodology \\comparison \end{tabular} 
& \begin{tabular}[c]{@{}l@{}} Geographical distribution, time \\zone, cultural diversity and physical\\ differences were reasons for \\miscommunication when conducting\\ requirements elicitation in \\GSD.\end{tabular} \\ \midrule

Khan and Akbar \cite{RN2215} & \begin{tabular}[c]{@{}l@{}} Motivation \end{tabular} 
& \begin{tabular}[c]{@{}l@{}} Global software \\development (GSD) \\-requirements change \\management
\end{tabular} 
& \begin{tabular}[c]{@{}l@{}} Academic-based \\- SLR\end{tabular} 
& \begin{tabular}[c]{@{}l@{}} Extracted 25 motivators in \\requirements change management\\ and developed taxonomies of \\identified  motivators.\end{tabular} \\ \midrule

Aldave et al. \cite{RN2969} & Creativity 
& \begin{tabular}[c]{@{}l@{}} Agile software \\development\\ -requirements elicitation
\end{tabular} 
& \begin{tabular}[c]{@{}l@{}} Academic-based \\- SLR\end{tabular} 
& \begin{tabular}[c]{@{}l@{}} Creativity brings innovations to\\ the project and can be successfully \\implemented in agile based \\software projects \end{tabular} \\ \midrule

Askarinejadamiri Z. \cite{RN2430} & Personality
& \begin{tabular}[c]{@{}l@{}} Web development \\ -requirements elicitation
\end{tabular} 
& \begin{tabular}[c]{@{}l@{}} Academic-based \\- SLR\end{tabular} 
& \begin{tabular}[c]{@{}l@{}} There is a relationship between \\personalities and RE in web \\development and need more\\ research considering its impact  \end{tabular} \\ \midrule

 Viller et al. \cite{RN1634} &  \begin{tabular}[c]{@{}l@{}} Human factors\\(Human errors) 
\end{tabular}
 & \begin{tabular}[c]{@{}l@{}} Dependable system \\development \\-RE
\end{tabular} 
& \begin{tabular}[c]{@{}l@{}} Academic-based \\- Survey on social \\and organisational\\ literature \end{tabular} 
& \begin{tabular}[c]{@{}l@{}} Identified various human errors\\ as an hierarchical classification \\according to the cognitive levels \\when people performing various\\ tasks (e.g., skilled-based, rule-based\\ and knowledge-based errors) \end{tabular} \\ \midrule

 Alsanoosy et al. \cite{RN1635} & Culture 
 & \begin{tabular}[c]{@{}l@{}} Requirements engineering\\ process - (Saudi Arabia \& \\Australian culture) \end{tabular} 
& \begin{tabular}[c]{@{}l@{}} Industry-based \\- interviews of\\ 16 RE practitioners \end{tabular} 
& \begin{tabular}[c]{@{}l@{}} Both Saudi arabian and Australian\\ cultures have an impact on RE and \\ provide various mitigation\\ strategies to overcome cultural\\ impact. \end{tabular} \\ \midrule

 Mendes et al. \cite{RN1602} & \begin{tabular}[c]{@{}l@{}} Personality and \\ decision-making style \\(DMS) \end{tabular}
 & \begin{tabular}[c]{@{}l@{}} Software project \\development \\ ( Brazilian context) \end{tabular} 
& \begin{tabular}[c]{@{}l@{}} Industry-based \\- survey with 63 Brazilian\\ software engineers \end{tabular} 
& \begin{tabular}[c]{@{}l@{}} Identified seven statistically \\significant correlations between \\DMS and personality, and\\ built a regression model considering\\ the DMS as the response variable\\ and personality factors as \\independent variables \end{tabular} \\ \midrule

Xia et al. \cite{RN3005} & \begin{tabular}[c]{@{}l@{}} Personality \end{tabular}
 & \begin{tabular}[c]{@{}l@{}} Software project \\development - project\\ success (2 large IT \\organisations in China)  \end{tabular} 
& \begin{tabular}[c]{@{}l@{}} Industry-based \\- survey with 346 \\software professionals \\of 28 teams\end{tabular} 
& \begin{tabular}[c]{@{}l@{}} Identified project manager \\personality and team personality\\ affect the success of software\\ projects and suggest to focus on\\ relationships between personality \\and SE activities\end{tabular} \\ \midrule

\begin{tabular}[c]{@{}l@{}}Fernandez-Sanz \\and Misra \cite{RN1636} \end{tabular} 
& \begin{tabular}[c]{@{}l@{}} Culture and Gender \end{tabular}
 & \begin{tabular}[c]{@{}l@{}} Team work performance \\ for software requirements \\analysis  (multinational \\environments)  \end{tabular} 
& \begin{tabular}[c]{@{}l@{}} Industry-based \\- case study with 35 \\ software professionals \end{tabular} 
& \begin{tabular}[c]{@{}l@{}} Identified differences depending on \\country or culture, but no significant \\differences related to gender (But \\require more female participants for \\solid conclusion)\end{tabular} \\ \midrule

\colorbox{lightgray}{\textbf{This Study (Hidellaarachchi et al.)}}
&  \colorbox{lightgray}{\textbf{\begin{tabular}[c]{@{}l@{}} List of human aspects \\ - Figure \ref{Fig 1: Human aspects in RE} \\ (specifically focusing \\motivation and \\personality) \end{tabular} }}
 & \colorbox{lightgray}{\textbf{\begin{tabular}[c]{@{}l@{}} RE-related activities \\in SE \\(Agile/ traditional\\/both)   \end{tabular} }}
& \colorbox{lightgray}{\textbf{\begin{tabular}[c]{@{}l@{}} Industry-based - \\survey with 111 \\software practitioners \\involved in RE\\ activities  \end{tabular} }}
& \colorbox{lightgray}{\textbf{\begin{tabular}[c]{@{}l@{}} Identified the practitioners' \\perspective on the level of \\influence of human aspects \\when involved in RE, a set \\of motivation factors that \\have an impact on RE, the \\importance of personality\\ aspect on RE and a set of\\ factors that make individuals \\less effective when involved\\ in RE-related activities\end{tabular}}} \\ \midrule

\end{tabular}%
}
\end{table*}

\par With these findings we were able to emphasize the contribution of our study compared with other existing work in the area. As shown in table \ref{Table 5: comparison with related work}, we have summarised a few of most related studies in our related area and compared it with our study to show how our contribution differs from other studies. There are number of studies that focused on various human aspects related to the RE context  where we have analysed them in our SLR \cite{RN1600}. With that we have identified what has been studied in this area and that the majority of the existing work are academic-based studies. There is no particular study that focused on identifying software practitioners perspective on the influence of various human aspects on RE-related activities. Hence, by conducting this study we wanted to present software practitioners' perspectives on the influence of various human aspects on RE-related activities that will be beneficial for future research studies.

\par As summarised in table \ref{Table 5: comparison with related work}, there are studies that have focused on various human aspects. However, they are mostly limited to focusing on only one or two human aspects and a specific context. For example, Coughlan and Macredie \cite{RN2952} focused on the impact of communication in requirements elicitation, but specific to global software development whereas Fernandez-Sanz and Misra \cite{RN1636} studied on culture and gender aspects on requirements analysis in multinational environments. This indicates that these studies either limit their focus to particular human aspect or RE-related activity. Hence, in our study, we have focused on obtaining participants' perspective on a list of human aspects with detail information on motivation factors, personality characteristics, less effective factors and important factors to measure performance which can be directed to in-depth future studies.

\subsection{Insights and Reflections} \label{section 5.2}
Based on our qualitative analysis of open-text answers and the memos we wrote when following the \textit{STGT for data analysis} approach, we have identified several interesting insights from our study. The following are the key insights and some reflections that we have obtained from our memos. These insights serve as suggestions for future work, while the reflections capture implications for researchers to consider.\\
 %\begin{itemize}
%\item \textit{\textcolor{black}{Insights on identified key motivation factors}}
%\end{itemize}

\par $\bullet$  \textbf{Importance of teamwork:}  It seems that our survey participants were highly concerned about team-related aspects such as \emph{teamwork, team collaboration, team behaviour} when it comes to both their motivation and reducing their effectiveness. They have described them with terms such as \emph{``cooperation among teams", ``better engagement with the team", ``collaborative work with the team"} and described the impact on RE/SE activities. Also, they linked them with other aspects such as ``\textit{team opinions}'', ``\textit{friendly environment}'' and ``\textit{empowered team}''. For example, R28 (business analyst) thinks by doing design sprints, he/she can connect with other team members and improve the collaborative teamwork. Also, R56 thinks a collaborative team is key to an empowered team that can take decisions on requirements. These responses highlight the importance that RE practitioners place on teamwork and how much they depend on their teams for their RE-related activities. Future studies can explore the role of teamwork in RE-related activities. 

%apart from their individual and team related aspects

\par $\bullet$  \textbf{Impact of external forces:}   We identified that there are various external forces that have an impact on practitioners as they are involved in RE-related activities. Among them, \textit{nature of the customers/clients} and \textit{management issues} play important roles. For example, it was mentioned that inexperienced customers, customers with unrealistic demands or lack of commitment make a significant impact when they perform RE-related activities. Also, management issues such as organisational politics and poor decisions from senior management were also mentioned by the participants that have an impact when involved in RE-related activities. These insights open avenues for future research into the role of customers and senior management in RE-related activities.

\par $\bullet$  \textbf{Inadequate requirements:}   Based on the evidence, the presence of inadequate requirements is another insight that emerged from our qualitative data analysis. For example, participants mentioned that \emph{unclear, unimportant} requirements and \textit{gaps} in requirements are aspects that should be paid attention to when involved in RE-related activities. They described that ``\emph{when customers didn't specify the requirements properly}'' or ``\emph{lack of understanding of what clients exactly need}'' make them less effective. Also, they mentioned issues of having only high-level requirements or not knowing the actual key customer needs and further explained their impact on RE and SE (Table \ref{TABLE 6: less effectiveness factors}). This points to a need to further investigate the impact of inadequate requirements and other barriers to effective performance in RE-related activities in future studies.\\

%However, we understand that there are vast variations of requirements in software development projects and identifying its nature and impact would be a challenging one. 

\noindent In addition to these insights, we also captured some key \textbf{reflections} based on our memoing:\\

\par $\bullet$  \textbf{Impact of subjective terms:} Based on participants' answers to our open-ended questions, we have noted the use of subjective terms to express ideas. For example, some respondents (n=11) considered customer/client/stakeholder engagement as a motivation factor. There, they mostly referred to  \emph{"good relationship or good bond"} with customers. But, what they actually mean by a ``\textit{good}'' relationship is not clear. For example, R46 (IT project manager) explicitly linked it to \emph{"frequent customer visits"} where R21 (senior business analyst) linked it with customer satisfaction. The use of subjective terms suggests the need for follow up clarifications and concrete examples to elucidate better meaning, emphasising the limitations of surveys and pointing to the need for semi-structured interviews as a means to collect detailed information about the key aspects.

\par $\bullet$ \textbf{Emergence of socio-technical concepts:} In terms of categorisation, it is really difficult to neatly group some of the factors into one particular group. For example, categorizing \emph{domain knowledge} was one of the more challenging parts of the analysis as it relates to both human and technical aspects. From our previous study \cite{RN1600}, we identified it as a "technical-related human aspect". Practically, domain knowledge can be best described as a socio-technical factor. However, when analysing open-text answers, we observed that participants tend to refer to it from the technical aspect (technical knowledge, product/project knowledge). Therefore, based on the evidence in this study, we grouped domain knowledge under technical aspect. Future studies can design data collection mechanisms to better capture the full socio-technical context of such concepts, e.g. by asking for examples of both the social and technical contexts of use, as well as contexts where the social and technical play out together and/or in tandem.

\subsection{\textbf{Recommendations}} \label{section 6}
Based on the findings from our survey, we have identified several key challenges in the area of human aspects impacting RE, which need more focus on. We have framed these as a set of recommendations, shared below, for software practitioners involved in RE-related activities, and the wider SE research community for further research into the human aspects of RE. 

%Various human aspects of the individuals involved in RE should be paid more attention to and incorporated when conducting RE-related activities

\textbf{1. Emphasising the role of human aspects in RE:} According to the responses of the software practitioners, various human aspects influence their performance when involved in RE-related activities. From the list of human aspects we provided based on our literature analysis \cite{RN1600}; the majority are considered to be highly important by the software practitioners (e.g., motivation, domain knowledge, attitude, communication skills and personality). Hence, it is important that these human aspects should be paid more attention to by software practitioners when involved in RE-related activities, and by SE/RE researchers to conduct more research on incorporating these aspects for the improvement of RE-related activities. For example, identifying a set of factors that motivate/de-motivate software practitioners when involved in RE-related activities would be beneficial for the software teams to improve the positive impacts while finding solutions to mitigate the negative impacts.

%Some are considered moderately important (e.g, emotions, cultural diversity, geographic distribution), whereas human aspects such as gender and age were considered to be the least important aspects.

%Importance of focusing the influence of various human aspects on the RE-related activities. .. Our survey focused on identifying the human aspects that are considered influential by software practitioners during RE-related activities. 

\textbf{2. Conducting in-depth studies to build a body of knowledge on human aspects in RE:} As one of our main focus area was \textit{motivation}, we were able to identify factors that positively and/or negatively influence motivation when involved in RE-related activities (see Figure \ref{Fig 6: Importance of Human aspects}), including the direction of the influence (positive and/or negative) and correlations. Similarly, future research should focus on other human aspects and conduct in-depth studies to collectively and over time, build a body of knowledge around human aspects in RE.

\par \textbf{3. Focusing on personality characteristics individually and as a team:}  The software practitioners surveyed agreed that personality characteristics of the people involved in RE-related activities are important in order to perform RE-related activities. Moreover, software practitioners considered that the majority of these characteristics listed in section \ref{section 4.4} are important to have within themselves as well as within their team members.  As these characteristics resemble individual personality traits, this suggests that the personality of the people involved in RE-related activities can be an important aspect and there is potential for further studies to focus on the influence of individual personality traits on RE-related activities.

%\textcolor{red}{Rashina says: Thus \#5 is a summary of the findings. There is no recommendation here! Remove.}

% \textbf{5. Correctness, completeness and clarity of requirements are highly important factors when measuring the performance of the individuals involved in RE-related activities: } 87.1\% of participants mentioned that all the given list of factors are extremely/very important in measuring the performance of the individuals involved in RE-related activities. Among them, the correctness of the requirements identified, the clarity of the requirements, and the completeness of the requirements identified were rated as the highest on measuring the individuals' performance. This implies that in practitioners perspective, these factors play an important role and can be considered when measuring performance of individuals who are involved in RE-related activities  (section \ref{section 4.6}).

\textbf{4. Better understanding the factors that make  individuals less effective when involved in RE-related activities: } We identified a set of factors that our participants think can make individuals less effective when involved in RE-related activities. We recommend that studies be carried out to study whether software practitioners having better understanding of these factors and their impacts enables them to actually be more effective (or not). Our survey findings confirm that communication issues, management issues, nature of the clients, behaviour of the team, unclear requirements, constant requirements changes, low domain knowledge, and lack of resources are all major factors that make individuals less effective at RE-related activities. As explained in Table \ref{TABLE 6: less effectiveness factors}, as these factors impact the  overall SE process, more attention should be paid to manage them in order to improve their impacts.

%Our survey reports respondents' opinions about these, and thus further studies are also needed to see if these opinions are the same as actual impacts on RE-related activities and overall SE project outcomes. [Rashina says: again undermining survey research, remove.]

\par \textbf{5. Investigating customer/client/end-user perspective on the influence of human aspects on RE-related activities:} The focus of our study was on software practitioners, specifically those involved in RE-related activities.  As RE-related activities also depend on the customers, clients, and end-users, we recommend also studying their perspectives on the influence of human aspects when involving RE-related activities.

% 8. Comparing the findings with more demographic details on participants:

%\textcolor{red}{Rashina says: I have rewritten \#8 to acknowledge reviewer's points about contextual factors, while trying not to undermine the contributions of our own work. Fix numbering in paper and response letter}

\par \textbf{6. Investigating contextual factors: } \textcolor{black}{Our survey study aimed to gain an overview of the key factors influencing practitioners when conducting RE-related activities, with a particular focus on motivation and personality. To this end, we collected some basic demographic and contextual information about the participants such as their educational levels, countries, and job roles. In addition to human aspects, a number of other contextual factors can impact RE-related activities. Contextual factors such as team size, organisation size, national and work cultures, the type of RE they are involved in (e.g. bespoke or market-driven) should also be investigated in future studies to understand their possible impact on human aspects, RE-related activities, and on the human aspects in the context of RE-related activities.}

%\textcolor{red}{Rashina says: I have added this new point. Currently numbering as 9, fix the numbering when you fix the other points.}

\par \textbf{7. Working toward a taxonomy of human aspects in RE: } From our SLR \cite{RN1600} and this survey study, it is clear that there are a number of human aspects that come into play when considering their impact on RE, e.g. age, gender, motivation, personality. However, there is no comprehensive and agreed taxonomy of human aspects in SE in general and RE in particular. Thus, there is a need to develop a taxonomy of human aspects that can be used to clearly define, study, and explain human aspects in RE.

\par \textcolor{black}{\textbf{8. Investigating impact of human aspects on RE work performance evaluation.} We wanted to see what factors are viewed as important when evaluating RE work performance in RQ5. However we received limited insights, especially regarding how human aspects may effect RE work performance assessment. Investigating how different human aspects may impact RE task performance appraisal would be very interesting and potentially important future research. }

% The software practitioners we surveyed belong to a variety of organizations in the software industry. The domain they work in, the type of RE-related activities they are involved in, and the team \& organization sizes and organisational cultures likely all differ. It would be valuable to find more information about how differing software domain, size of the organization and their teams, and the frequency of their involvement in RE-related activities  may impact our findings.} %However, there is a feasibility issue of considering all these aspects in one study and hence, we recommend considering these aspects in future studies that we have already incorporated in our next studies.}

\section{\textbf{Limitations and Threats to Validity}} \label{section 7}
Considering threats to \textbf{external validity}, though we shared the survey on social media and AMT aiming to obtain worldwide participation, we could not achieve it and found that the majority of the participants (42\%) were from Sri Lanka (Table \ref{TABLE 1: Participants' demographics}). Hence similar to the study \cite{RN1610}, our findings may also be biased and limit generalizing to the entire global software engineering community. However, in practice, such generalization is unlikely achievable. Considering participants' job roles, there were 17 different job roles with the rage of less than 1 year to more than 10 years of experience in software industry. However, different organisations may have different interpretations to these job roles which may need detailed explanations of each job role. To mitigate this (as our focus was to obtain data from software practitioners involved in RE-related activities), we provided a set key RE-related job responsibilities and asked them to rate their involvement ("Never" to "Always") to see whether they are actually involved in RE-related activities. We asked them to provide any additional tasks within their job roles. 34 of 111 mentioned testing user requirements, designing the solution/prototyping, and team and customer management as other job responsibilities which indicates that we achieved collecting data from our target participants.  \textcolor{black}{Since our main focus in this work was to gain understanding about participants' perspectives on the influence of human aspects on RE-related activities, specifically focusing motivation and personality aspects, we have not considered their team or organisation size details, software domain, or details about the frequency or amount of doing RE-related activities, which is a limitation for this study. Hence, the participants of this study may be involved in various types of RE in various contexts such as bespoke or market-driven projects and the limitation of these contextual factors may have an impact on our findings. These thus need to be further validated in future studies considering these factors (see section \ref{section 6} - recommendation 6).  To make our survey tractable to practitioners we included some basic demographic questions and opted to investigate more detailed demographic and work contexts in future studies.  }

Among threats to \textbf{internal validity}, we identified that using a survey as a research tool could be a limitation with regards to detailed data collection. But, since we wanted to obtain data from large number of software practitioners worldwide, a survey would be the most suitable approach. Additionally, with the global pandemic causing extended lockdowns in Melbourne, we were severely restricted in the possibility of conducting interviews. Also, we had to consider the time taken to complete the survey as our target participants were from industry and as a result we could only ask limited set of questions which can be completed within a feasible time period for them. Hence, we limited our survey to the most important questions that we needed to ask considering our future research studies. For example, we limited the open-ended questions only focusing on the influence of motivation aspect and other human aspects were mentioned only in a closed-ended questions to rate their importance just to get their initial idea on it.

% Though the survey was designed with both closed and open-ended questions, some participants have not provided complete answers to the open-ended questions. We got some short responses that were not clear or usable in our analysis. For example, R44 (Business Analyst), mentioned \emph{"conflicts"} as a factor that makes him/her less effective. But it was not clear whether he/she referred to conflicts with the team or clients, conflicts on what, which was another practical issue when analysing open-ended questions in surveys as we were unable to clarify it.
Also, when following the STGT for qualitative data analysis, generating some concepts and categories based on the codes were a challenge for survey responses compared with in-depth interview responses.  However, we selected STGT approach for our qualitative data analysis as it leads to original, relevant and in-depth findings as well as valuable insights for future studies through its reflective memo writing practice. For example, when the concepts were grouped into two categories named collaborative \& human aspects and technical aspects when identifying motivation factors, grouping the concepts such as domain knowledge was a challenge. From our previous study \cite{RN1600}, we identified that there is a need of a more comprehensive and agreed taxonomy of human aspects in SE in general and hence, we have grouped them considering its definitions used in SE studies. There, when grouping the human aspects, the domain knowledge was grouped as a "technical-related human aspect" where we used it under the list of human aspects for this study. However, when identifying motivation factors, we found out that there are some human aspects that act as motivation factors as well, including domain knowledge. Considering its practicability, domain knowledge is more likely to identify as a socio-technical aspect. However, following the limited open-text answers, we have grouped it into the technical aspect category, which is the best match category with the available data. Hence, as explained in section \ref{section 4.3}, we have considered domain knowledge as a human aspect as well as a motivation factor according to the participants perspective. 
\textcolor{black}{Participants may interpret some of the terms used in our survey in different ways than we intended. For example, `performance' might be interpreted by one developer assuming it means writing many well-formulated requirements, while another person may interpret performance as writing ``correct" requirements. We tried to mitigate this by using factors identified in the literature shown to impact RE task performance and an open ended question to explore further the respondent's interpretation of performance. However, respondents may still have interpreted meanings of some terms in different ways.
}

\par Considering the number of participants of the survey, we identified that, although 207 participants started doing the survey (as shown in qualtrics records) but only 118 participants completed it fully. The target participants of the survey was software practitioners working in the industry. Due to that, we had to remove 7 participants who have completed the survey as they were students (undergraduates/postgraduates) or IT teachers or lecturers. As a result, we had to consider only the answers of the remaining 111 participants who completed the survey. All the authors involved in each stage of the survey and in the analysis stage, for the quantitative data, the first author conducted the initial analysis and shared with others to discuss each followed steps and techniques. Based the discussions, the team finalized the best ways to present the findings and for the qualitative study, as explained in section \ref{section 3.3}, we conducted sessions on STGT approach for qualitative data analysis and had several discussions on the analysis, findings and ways to present the findings to mitigate the bias. Using payment for second round data collection can also be a threat to the internal validity of the research. However, we have decided to use AMT after careful considerations of similar studies \cite{RN1611}, and only approved the payments for the participants after examining their responses to check whether they belongs to our target participant group and if only they provided responses for each question. 

\newpage
\section{\textbf{Conclusion}} \label{section 9}
This investigation contributes to understanding the industry perspective on the influence of human related aspects when involved in RE-related activities. Our study results show that software practitioners greatly agree that the majority of human aspects provided based on the literature are important and influential based on their experience. Among them, \emph{motivation}, \emph{domain knowledge}, \emph{attitude}, \emph{communication skills}, and \emph{personality} were identified as highly important aspects. \emph{Emotions}, \emph{cultural diversity}, \emph{geographical distribution} were identified as moderately important. \emph{Gender} and \emph{age} were mentioned as the least important aspects. We focused on further investigations on motivation and personality aspects, and have identified a set of factors that motivate individuals when involved in RE-related activities and their impact on RE/SE activities. By analysing participants open-text answers, we have categorized these motivation factors into two major groups: collaborative \& human aspects and technical aspects. Among them, the majority were collaborative \& human aspects, and the impacts were mentioned related to either RE or SE in general. Among collaborative \& human aspects, \emph{team collaboration}, \emph{personal satisfaction}, \emph{customer/stakeholder engagement}, \emph{communication skills}, \emph{individual's interest}, \emph{customer satisfaction}, \emph{experience of individuals}, \emph{job satisfaction} were some of the commonly identified motivation factors, whereas some were specifically mentioned by one or two participants. Considering technical aspects, commonly identified motivation factors were \emph{domain knowledge}, \emph{clarity of requirements}, \emph{comprehensive RE}, \emph{accessibility to the resources}, and \emph{good outcomes}, whereas the \emph{complexity of the requirements}, \emph{deadlines}, \emph{documentation skills}, and \emph{design methods} were mentioned by one or two participants.

\par We have identified the perceived importance of some specific individual personality characteristics when conducting RE-related activities. Our participant software practitioners consider it is important to have these personality characteristics themselves as well as their team members. The results of our study also identified some factors that practitioners think make them less effective when involved in RE-related activities.  These factors include \emph{communication issues}, \emph{unclear requirements}, \emph{less domain knowledge} and \emph{lack of resources}. Other factors that make the individuals less effective include \emph{management issues}, \emph{nature of the clients/stakeholders}, \emph{team behaviour}, \emph{constant requirements changes} and \emph{issues in the design and the method}.   Our participants also shared some approaches to measuring the performance of individuals when involved in RE-related activities. These include the correctness of the requirements identified, clarity of the requirements, and completeness of the requirements. They think that these should be given the highest importance when measuring the performance when involved in RE-related activities. The findings of this study will be beneficial for understanding the industry perspective of the influence of various human aspects on RE-related activities, specifically related to motivation and personality aspects. The research community can get an idea of what software practitioners think and conduct studies that will benefit the industry. Meanwhile, software practitioners can take due consideration of these findings, when forming \& managing teams, and when conducting their RE-related activities.

\section*{\textbf{Acknowledgments}}
This work is supported by Monash Faculty of IT PhD scholarships. Grundy is supported by ARC Laureate Fellowship FL190100035 and this work is also partially supported by ARC Discovery Project DP200100020.



\bibliographystyle{ACM-Reference-Format}
\bibliography{References}

\appendix
\section{Survey Questions} \label{A}
\begin{footnotesize}
\textbf{Section 01: Personal Information}
\begin{enumerate}
  \item How old are you?
  \item How would you describe your gender?

\begin{tasks}[style=itemize, column-sep=-5mm, label-align=left, label-offset={0mm}, label-width={3mm}, item-indent={10mm}](4)%
\task Male
\task Female
\task Prefer to self-describe:
\task Prefer not to answer

\end{tasks}
   
  \item Country of your residence?
  \item Educational background: (choose one option)
  \begin{itemize}
      \item University degree in software engineering/ computer science
      \item University degree in other IT fields: 
      \item Associated degree/ diploma in software engineering/ computer science
      \item Associated degree/ diploma in other IT fields: 
      \item Other university degree/ associate degree/ diploma
  \end{itemize}
\end{enumerate}
\textbf{Section 02: Employment Information} 
  \begin{enumerate}
  \item What is your current job role/ job title?
  \item What type of software development methods you have majorly involved in? (select all that apply)
   \begin{tasks}[style=itemize, column-sep=-15mm, label-align=left, label-offset={0mm}, label-width={3mm}, item-indent={5mm}](3)%
\task Traditional (waterfall)
\task Agile (please specify):
\task Other (please specify):
\end{tasks}
  
  \item Your current job responsibilities include: (please rate the following based on your involvement): close-ended question with the likert scale  from \emph{"Never"} to \emph{"Always"}.
  \begin{itemize}
      \item Collaborate with the stakeholders to elicit requirements
      \item Documenting software requirements specifications according to standard templates
      \item Lead requirements analysis and verification
      \item Participate in requirements prioritization
      \item Manage requirements throughout the project
  \end{itemize}
  
  \item If you think there are any other job responsibilities that you involved in apart from the above list, please specify them:
  \item Your experience in carrying out requirements engineering related tasks:
 \begin{tasks}[style=itemize, column-sep=-15mm, label-align=left, label-offset={0mm}, label-width={3mm}, item-indent={5mm}](3)%
\task No experience 
\task Less than 1 year
\task Between 1 to 5 years
\task Between 6 to 10 years
\task  More than 10 years
\end{tasks} 
 
 \end{enumerate} 
\textbf{Section 03: Performance in Requirements Engineering}
  \begin{enumerate}
\item How important are the following factors in measuring the performance of the people involved in Requirements Engineering? (Please choose one option for each factor): close-ended question with the likert scale from \emph{"Not at all important"} to \emph{"Extremely important"}.

\begin{itemize}
    \item The correctness of the requirements identified
    \item Completeness of the requirements identified
    \item Clarity of the identified requirements
    \item Ability to respond to the requirements changes
    \item Ability to incorporate customer feedback
    \item Ability to incorporate software team feedback
    \item Ability to interact with others involved in requirements engineering
\end{itemize}
 \item Are there any other factor(s) apart from the above list that you use to measure individual performance in requirements engineering activities, please mention: (If not, please mention "none")
\item Please indicate to what degree you agree or disagree with the following statements\\
\emph{"The success of the requirements engineering greatly depends on the people involved in the requirements engineering activities as their performance varies from one another"}
 
\begin{tasks}[style=itemize, column-sep=-15mm, label-align=left, label-offset={0mm}, label-width={3mm}, item-indent={5mm}](3)%
\task Strongly agree
\task Agree
\task Somewhat agree
\task Neither agree nor disagree
\task Somewhat disagree
\task Disagree
\task Strongly disagree
\end{tasks}
 
 \emph{"Differences in characteristics, behaviours, personal habits, skills of the people involved in the requirements engineering activities affect the requirements engineering process"}   
\begin{tasks}[style=itemize, column-sep=-15mm, label-align=left, label-offset={0mm}, label-width={3mm}, item-indent={5mm}](3)%
\task Strongly agree
\task Agree
\task Somewhat agree
\task Neither agree nor disagree
\task Somewhat disagree
\task Disagree
\task Strongly disagree
\end{tasks}
    
    \item In your experience, to what extent do the following human aspects influence the performance of the people involved in requirements engineering? (Please choose one option for each factor): close-ended question with the likert scale from \emph{"Not at all important"} to \emph{"Extremely important"}.
    \begin{tasks}[style=itemize, column-sep=-15mm, label-align=left, label-offset={0mm}, label-width={3mm}, item-indent={5mm}](4)%
\task Personality
\task Motivation
\task Gender
\task Cultural diversity
\task Emotions
\task Domain knowledge
\task Attitude
\task Age
\task Communication skills
\task Geographic distribution
\end{tasks}

    \item If there are any other human aspect(s) apart from the above list that influences the performance of the people involved in requirements engineering activities, please mention:
    \item In your opinion, what are the factors that motivate you to perform effectively in requirements engineering activities? Please explain briefly why. 
    \item In your opinion, what are the factors that make you ineffective when involved in requirements engineering? Please explain briefly why. 
\end{enumerate}
\textbf{Section 04: Characteristics of the people involved in the RE process}
\begin{enumerate}
    \item Please rate this question based on which of your personal human characteristics are important to conduct requirements engineering activities effectively. (Please choose one option for each factor):  close-ended question with the likert scale from \emph{"Not at all important"} to \emph{"Extremely important"}.
    \begin{itemize}
        \item Enjoy interacting with people
        \item Are enthusiastic about what they do
        \item Display intellectual curiosity
        \item Are willing to try new things
        \item Strive for high achievements
        \item Prefer following a plan over spontaneous behaviour
        \item Have kind, generous, trustworthy, helpful qualities
        \item Are willing to compromise
        \item Have  a tendency towards negative emotions
        \item Get stressed out easily
    \end{itemize}
      \item Please rate this question based on which of your team members human characteristics are important to conduct requirements engineering activities effectively. (Please choose one option for each factor): close-ended question with the likert scale from \emph{"Not at all important"} to \emph{"Extremely important"}.
    \begin{itemize}
        \item Enjoy interacting with people
        \item Are enthusiastic about what they do
        \item Display intellectual curiosity
        \item Are willing to try new things
        \item Strive for high achievements
        \item Prefer following a plan over spontaneous behaviour
        \item Have kind, generous, trustworthy, helpful qualities
        \item Are willing to compromise
        \item Have  a tendency towards negative emotions
        \item Get stressed out easily
    \end{itemize}
      \item If you think there are any other key human characteristics that are important for the people involved in the requirements engineering process, please specify them;
      \item What are the requirements engineering activities that you have considered when answering to the above question? (eg: requirements elicitation, analysis, specification, validation, management)
      \item Are you willing to participate in a personality test in our next phase of the project to identify your personality traits that may influence the performance on the requirements engineering activities? (Yes/ No)
      \item Please provide the following information to contact you for the personality test in future: (Name/ Email address)
      \item Any other feedback on this survey?
      
\end{enumerate}

\end{footnotesize}
\end{document}